\documentclass[10pt, twocolumn]{IEEEtran}
\usepackage[utf8]{inputenc}
\usepackage{lmodern}
\usepackage{newtxtext}
\usepackage{mathtools, amssymb, bm, bbm}
\usepackage{dsfont}
\usepackage{xcolor, cite}
\usepackage{url, hyperref}
\usepackage{subfigure}
\usepackage{float}
\usepackage[utf8]{inputenc}

\newcommand{\h}{\mathbbm{h} }

\newcommand{\E}{\mathbbm{E} }

\DeclareMathOperator*{\argmax}{arg\,max}

\usepackage{setspace}
\allowdisplaybreaks

\title{ 3-D Position Optimization of Solar-Powered Hovering UAV Relay in Optical Wireless Backhaul  }
\author{Heyou Liu, Muhammad Salman Bashir, \IEEEmembership{Senior Member, IEEE}, Mohamed-Slim~Alouini,~\IEEEmembership{Fellow,~IEEE} \thanks{H.~Liu and M.-S.~Alouini are with the CEMSE Division, King Abdullah University of Science and Technology (KAUST), Thuwal 23955-6900, Kingdom of Saudi Arabia. M.~S.~Bashir is with the School of Computing and Engineering, University of Huddersfield, Huddersfield HD1 3DH United Kingdom. e-mail: (heyou.liu@kaust.edu.sa; m.bashir@hud.ac.uk; slim.alouini@kaust.edu.sa)}}
\date{December 2021}

\begin{document}

\maketitle

\begin{abstract}
A major hurdle in widespread deployment of UAVs (unmanned aerial vehicle) in existing communications infrastructure is the limited UAV onboard energy. Therefore, this study considers solar energy harvesting UAVs for wireless communications. In this context, we consider three dimensional position optimization of a solar-powered UAV relay that connects a distant sensor field to an optical ground station (OGS) for data processing. The integrated sensor-UAV-OGS network utilizes radio frequency band for sensor-to-UAV links and the optical band for the UAV-to-OGS feeder link.  Since atmospheric conditions affect both the harvested solar energy as well as the optical wireless signal, this study tackles UAV position optimization problems under  various channel conditions such as clouds, atmospheric turbulence and dirt. From this study, we discover that the optimum position of the UAV---that maximizes the end-to-end channel capacity---is heavily dependent on the atmospheric channel conditions.  
\end{abstract}


\section{Introduction}

Unmanned Aerial Vehicles (UAVs) are revolutionizing traditional paradigms in contemporary wireless communication. Their unique ability to navigate difficult terrains and regions grants UAVs a substantial advantage in terms of deployment flexibility, cost efficiency, and adaptability over static relaying methods \cite{Abeywickrama:18:Access, 7572068}. This is particularly advantageous in areas without ground infrastructure or where ground terminals are separated by obstacles such as mountains or buildings as UAVs can essentially provide  line-of-sight (LoS) connectivity and enhanced capacity \cite{lyu2016placement,10024838}. 
 The mobility of UAVs is ideal for on-demand operations and can significantly improve communication channels, primarily through the establishment of short-distance LoS links \cite{qin2023downlink, zeng2017energy}. Utilized as aerial base stations, UAVs are particularly beneficial in scenarios that demand immediate and adaptable wireless service, such as in temporary hotspots and emergency situations. The enhanced availability of LoS connections due to UAVs' elevated positions, combined with their high maneuverability, allows for a fast and highly flexible deployment of communication infrastructure in time-sensitive or economically restrictive environments \cite{7486987, sun2019optimal}. 

 \subsection{Problem Motivation} \label{motivation_section}
Despite their extensive applications, the deployment of UAVs in communication systems encounters significant challenges, predominantly related to onboard energy storage limitations. The UAVs' endurance and performance are directly restricted by the finite energy reserves due to  size and weight constraints of aerial platforms \cite{kandeepan2014aerial, 7192644, Chu:21:Drones}. This necessitates a strategic focus on energy-efficient communications and exploitation of energy harvesting techniques to maximize the operational time of the UAV. Two popular energy harvesting techniques for optical wireless based UAVs are i) \emph{simultaneous lightwave information and power transfer} (SLIPT) in which the data carrying laser beam from the ground is also used to power the UAV \cite{Diamantoulakis:18, Diamantoulakis:TGCN:18, Bashir:TWC:22} and ii) solar-powered UAVs that harvest  sunlight energy to either partially or fully meet their power requirements \cite{7500855, 7353711, sun2019optimal}.

There are only a few studies that discuss the trajectory/position optimization of solar-powered UAVs to maximize some quality-of-service metric (such as channel capacity) in wireless communications. Specifically, articles \cite{sun2019optimal, 9369086} thoroughly discuss the resource allocation and trajectory optimization of a solar-powered UAV to maximize the sum rate in the multi-user and multi-sensor links, respectively. Both these studies consider RF channels to communicate with users/sensors. 
However, for serial RF/FSO links in which the UAV connects the multi-user (or multi-sensor) side to a optical ground station (OGS) through i) RF (user-to-UAV) links and ii) a feeder FSO (UAV-to-OGS) link, position optimization of a UAV relay incurs extra considerations which have not been addressed before to the best of our understanding. Firstly, for the two-dimensional UAV position optimization (for UAVs obtaining sufficient sunlight even at lower heights), the end-to-end rate is limited not only by the user-to-UAV RF link but also the UAV-to-OGS optical link. Secondly, for UAVs limited by solar energy, atmospheric effects---such as clouds and dirt---will play a major role not only in affecting the harvested solar energy but also the FSO laser signal connecting the UAV and the OGS. Thirdly, for the solar energy limited UAV, UAV position optimization incurs certain tradeoffs between maximizing harvested solar energy and the end-to-end rate of the serial RF/FSO link. All these considerations form the motivation of the current study which deals with the optimization of a solar-powered UAV relay position in an  RF/FSO communication system.

\subsection{Background Literature Review}

Recent literature on energy efficient UAV-based wireless communication systems considers research efforts primarily focused on two interrelated objectives: energy resources optimization and UAV placement (or UAV trajectory) strategies to enhance system performance. A core aspect involves the UAVs' role as either aerial base stations, or relays, to enhance support for ground users \cite{bor2016efficient}. Trajectory optimization emerges as a key question in the study \cite{9217992} which optimizes UAV relay paths to maximize ground terminal throughput and improve static relaying methods. In \cite{zeng2017energy}, trajectory optimization also accounts for energy efficiency by considering propulsion energy consumption alongside communication throughput. Articles \cite{fan2018optimal, 7572068} collectively examine joint optimization of transmission power, bandwidth, and UAV positioning to improve the overall performance of UAV-assisted networks.  The works  \cite{yang2018joint, Ye:20:WCL} discuss power consumption concerns which focus on minimizing uplink power for ground users and exploring energy-constrained operational strategies, respectively.

Limited onboard energy has forced the UAV design in recent years to focus on an energy-efficient paradigm.  To address the finite battery life, which is a critical limitation for UAVs, some studies such as \cite{wu2018joint, 7486987, 7572068, Abeywickrama:18:Access} have proposed comprehensive energy consumption models that take into account various aspects and functions of UAVs. However, reliance on onboard batteries for UAV-based communication systems leads to  limited operation times. The frequent need for these UAVs to return to base for recharging hinders the stability and sustainability of communication services, thus creating system performance bottlenecks. 
Although direct battery replacement or recharging is a straightforward approach to extend operational life, it becomes impractical and labor-intensive when scaling up to a large number of UAVs or sensors \cite{Ye:20:WCL}. Thus, seeking innovative methods to bolster the energy framework of UAVs to support extended and reliable operations continues to be a critical challenge in the design and deployment of UAV communication systems.

To overcome the limitations of conventional battery systems due to finite on-board energy, the potential of solar-powered UAVs has been explored in recent literature. The authors in \cite{7500855, 7353711} described the development of solar-powered UAV prototypes that leverage solar panels to  efficiently convert solar energy into electrical power to enable extended flight duration. The authors in \cite{9501239} introduce the idea and implementation of solar-powered UAVs as base stations for 5G and beyond systems.
The concept of solar-powered UAVs is further exemplified by prototypes  that have successfully demonstrated long-endurance flights exceeding a full day \cite{Chu:21:Drones, sun2019optimal}. This technology has garnered interest for its implications in both military and civilian spheres, particularly for tasks requiring prolonged aerial presence such as reconnaissance and surveillance \cite{7859311}. 
The advancement in solar cell materials technology, specifically the use of flexible and lightweight cells, is pivotal for the integration of solar energy into UAV systems. The authors in \cite{Zhang:18:Solar_Power} explored solar cells materials characterized by their nanocrystalline structure, robust light absorption across a wide spectrum and suitability for deployment on flexible substrates. These materials are pivotal for crafting ultra-lightweight and adaptable solar cells that can be seamlessly integrated into UAV designs to enhance their operational efficiency and expanding their utility in both aerial and terrestrial domains. Furthermore, the exploration of solar power extends beyond UAVs to other realms of transportation such as the six-seat transporter flight referenced in \cite{WinNT}. These innovations in solar cell materials and construction techniques have enhanced the practicality of solar-powered UAVs and broaden their application spectrum. 

This work is motivated by the idea that the efficacy of solar energy harvesting in UAV applications is closely intertwined with the UAV's operational altitude, as discussed in \cite{sun2019optimal, duffie2013solar, 7859311}. Atmospheric transmittance, a critical determinant of the solar energy captured by UAV-mounted photovoltaic cells, diminishes at lower altitudes due to increased air density and potential obstructions such as clouds. This attenuation of solar irradiance at reduced altitudes consequently affects the solar energy flux available for conversion to electrical power.
Conversely, as UAVs ascend to higher altitudes, they benefit from increased atmospheric transmittance, thereby enhancing the solar energy harvest. Yet, this advantage is not without its trade-offs. Elevated altitudes, while favorable for solar energy collection, concurrently escalate the path loss in air-to-ground communication links, presenting a trade-off that necessitates a balanced approach to optimize both solar energy harvesting and communications performance \cite{sun2019optimal,  9369086}. Therefore, the design of UAV systems requires a careful consideration of these conflicting aspects to ensure an optimal balance between energy harvesting and communication quality. In addition, the authors in \cite{7486987} and \cite{bor2016efficient} investigated the strategic 3-D deployment of UAVs, targeting optimal user coverage and service area expansion, with the former focusing on maximizing network revenue, and the latter on the deployment of UAVs as aerial base stations with directional antennas. Besides, the optimization of solar energy capture through intelligent path design is detailed in \cite{dai2012optimal} which employs quaternion-based methods, whereas the work in \cite{7859311} improves solar-powered UAV endurance by optimizing flight paths to exploit gravitational potential and attitude adjustments. The joint optimization of 3-D positioning trajectory and resource allocation is analyzed in studies \cite{7486987,7510820}, to enhance system throughput and extend UAV flight endurance. 
Expanding on these endeavors, the authors in \cite{Zhang:21:TMC} and \cite{Song:21:ICC} presented energy-efficient trajectory schemes that merge solar energy harvesting with charging stations, thereby constructing a sustainable communication service framework for UAV-assisted networks.  The comprehensive approach in \cite{sun2019optimal} incorporates trajectory planning with resource allocation for multi-carrier systems to maximize operational efficiency over extended periods. {Moreover, the authors in  \cite{9107998} developed a greedy algorithm augmented with a buffering mechanism to compute optimal flight trajectories that enables perpetual flight in solar-powered aircraft.}
In a similar vein, the work  \cite{9300182} presented a mission-oriented design approach for three-dimensional (3D) path planning of solar-powered unmanned aerial vehicles. This approach strategically utilizes limited solar energy resources to optimize the effectiveness of the mission. The article  \cite{Ye:20:WCL} explored a solar-powered UAV that is tasked with data collection from Internet of Things Devices (IoTDs) and providing energy to them through laser. The main focus of this article is on maximizing the UAV's residual energy by jointly optimizing the IoTDs' scheduling and the UAV's three-dimensional trajectory. An innovative iterative algorithm---that integrates block coordinate descent and successive convex optimization techniques---is recommended to efficiently address the simultaneous optimization of user scheduling and UAV path planning.

Overall, these references underscore the critical goal of achieving optimal UAV performance through a focus on solar power utilization and 3-D positioning. Thus, it is essential to formulate an integrated optimization strategy that encapsulates both altitude-dependent solar energy harvesting and the end-to-end performance of communication system under different environment conditions. This strategy should account for the UAV's flight trajectory or position, thereby ensuring optimal energy efficiency, maximum data transmission rates, and 
reliable communication.

\subsection{Contributions of Current Study}
The main contributions of this paper can be summarized as follows:
\begin{enumerate}

    \item The first contribution is the derivation of (approximate) closed-form solutions for optimal UAV position in two dimensions for the aggregate sensor-to-UAV channel (Section~\ref{2-D optimization}). These approximate solutions are highly accurate in asymptotic regimes. Furthermore, for the aggregate channel, we present a novel closed-form solution to determine the UAV's optimal coordinates within a max-min fairness framework.

    \item For the end-to-end link, the role of a solar-powered UAV as an amplify-and-forward (AF) and decode-and-forward (DF) relay is explored (Section~\ref{end-to-end optimization}). A number of asymptotic scenarios were covered for such relays which gave us deeper insights and helped us explain the optimization results pertaining to the optimum altitude of a solar-powered UAV relay.
    
    \item For the end-to-end link that comprises sensor-to-UAV and UAV-to-OGS links, we assessed the impacts of atmospheric turbulence and cloud layers on laser signal attenuation as well as the harvested solar energy, and explored their influence on the UAV's optimal positioning (Section~\ref{3-D optimization}). The criterion for optimal UAV positioning is the end-to-end channel capacity that is maximized for the UAV's role as both an amplify-and-forward (AF) and decode-and-forward (DF) relay. Various insights for optimal 2-D and 3-D positioning of UAV relay---based on atmospheric conditions---are developed.
\end{enumerate}

\subsection{Model Assumptions}
To simplify the analysis, we now consider the following two major  assumptions (along with justifications) in our study.
\begin{enumerate}
    \item We assume that the uplink RF channel between each sensor and the UAV is dominated by the LoS component. This assumption holds well since the beamwidth (in meters) from the transmitting sensor is small close to the ground and most sensors are ``looking upwards'' to the UAV while transmitting. This arrangement implies that the reflections from the ground objects are at a minimum, and therefore, the assumption of the dominance of LoS holds. Additionally, for the sake of simplicity, we assume that the RF signal experiences nominal fading in the LoS channel.
    \item The optical wireless feeder link from the UAV-to-OGS is modeled by the (signal independent) Gaussian noise at the OGS receiver. This assumption holds under moderate signal strength conditions at the OGS receiver. Here, we have used the assumption that the received signal is neither too large so that the signal-dependent shot noise model applies, nor the signal strength is so weak that the Poisson model is justified.
\end{enumerate}

\begin{figure}
    \centering
    \includegraphics[width = 0.8\columnwidth]{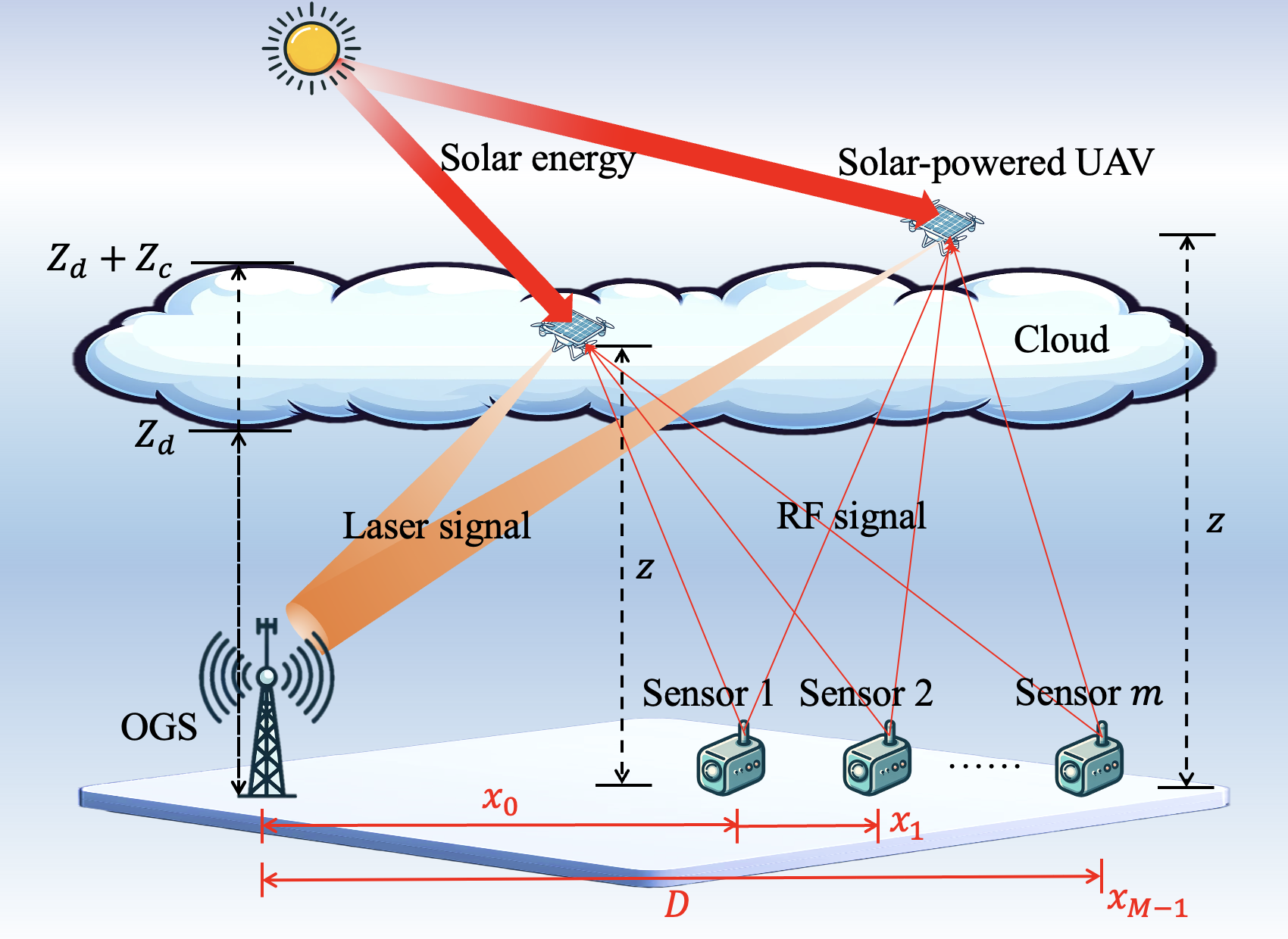}
    \vspace{-2mm}
     \caption{This figure depicts the system model of a  UAV-assisted serial RF/FSO link. This figure shows two scenarios where, in the first scenario, the UAV lies inside the cloud, and in the other scenario, the UAV hovers on top of the clouds.} 
    \label{Model}
\end{figure}

\subsection{Paper Organization}
This paper is organized as follows. Section~\ref{channel_model} presents the composite optical channel model in the UAV-to-OGS backhaul link. This is followed by Section~\ref{2-D optimization} that considers two-dimensional position optimization (in the $(x, y)$ plane) of a UAV relay for the RF based aggregate sensor-to-UAV link. The results obtained in this section give important insights for maximization of end-to-end rate of the serial RF/FSO link. Section~\ref{energy harvesting models} focuses on solar energy harvesting and energy consumption models of a hovering UAV. The results obtained in both Section~\ref{2-D optimization} and Section~\ref{energy harvesting models} are utilized in maximization of end-to-end rate in Section~\ref{end-to-end optimization}. The three-dimensional position optimization of a solar-powered UAV relay, along with a commentary on experimental results, is analyzed in Section~\ref{3-D optimization}. The main results of this study are summarized and concluded in Section~\ref{conc}.

\section{Backhaul Optical Signal Fading Models} \label{channel_model}

The (composite) channel gain of the UAV-OGS optical wireless link is given by
\begin{align}
    \h = \h_p  \h_s  h_c h_a,
\end{align}
where $\h_p$ represents the loss in energy due to pointing error and $\h_s$ corresponds to signal fading due to scintillation of the optical signal caused by atmospheric turbulence. The deterministic coefficients, $h_c$ and $h_a$, denote the attenuation of the backhaul channel due to (scattering and absorption) through cloud and air, respectively.

\vspace{-4mm}
\subsection{Pointing Error Model}
For a circularly symmetric Gaussian pointing error distribution in two dimensions, the magnitude of the error, $R$,  is distributed as a Rayleigh random variable. In this case, the random channel coefficient is \cite{Bashir:TCOMM:22}
\begin{align}
    \h_p \approx \frac{a_d^2}{2\theta^2(x^2 + y^2 + z^2)}\exp\left(-\frac{R^2}{2\theta^2(x^2+y^2+z^2)} \right),
\end{align}

and the distribution of $\h_p$ is \cite{Bashir:TWC:22} 
\begin{align}
    f_{\h_p}(h) &= \Phi h^{{(\theta^2 - \sigma^2)}/{\sigma^2}} \cdot \mathbbm{1}_{[0, B)} (h), \label{hrd}
\end{align}
where $\theta$ and $\sigma$ are the angular beamwidth and  angular pointing error standard deviation, respectively. Assuming that the OGS lies at the origin $(0, 0, 0)$, the point $(x, y, z)$ represents the location coordinate of the UAV relay, $\mathbbm{1}_A$ is the indicator function over a measurable set $A$, and
\begin{align}
\Phi & \coloneqq \frac{\theta^2}{\sigma^2} \left( \frac{1}{B} \right)^{{\theta^2}/{\sigma^2}}, \label{Phi} \\
B & \coloneqq  \frac{a_d^2}{2\theta^2(x^2 + y^2 + z^2) 
}. \label{B_1_val}
\end{align}
The factor $a_d$  in \eqref{B_1_val} is the radius of OGS receiver telescope.
\vspace{-4mm}
\subsection{Atmospheric Attenuation for Cloud and Air Model}

 We assume that the cloud layer is parallel to the ground and that both the upper and lower boundaries of the cloud layer remain parallel to each other (as shown in Fig.~\ref{Model}). Then, by the principles of similar triangles, we can deduce the link distance, $d_c$, through the cloud as
\begin{align}
    &\frac{Z_c}{z} = \frac{d_c}{\sqrt{x^2+y^2+z^2}}, \ z>Z_c+Z_d \nonumber \\
    &\frac{z-Z_d}{z} = \frac{d_c}{\sqrt{x^2+y^2+z^2}}, \ Z_d < z < Z_c+Z_d \nonumber \\
    \implies &d_c = \begin{cases}
        \frac{Z_c}{z} \sqrt{x^2+y^2+z^2}, &  z>Z_c+Z_d  \\ \frac{z-Z_d}{z}\sqrt{x^2+y^2+z^2}, & Z_d < z < Z_c + Z_d \\ 0, & z < Z_d,
    \end{cases}
\end{align}
where $Z_c$ is the thickness of cloud and $Z_d$ is the distance from the ground to the lower cloud boundary.  Thus, the link distance during the air is $d_a = \sqrt{x^2+y^2+z^2} - d_c$. Then, the values of channel coefficients, $\h_c$ and $\h_a$, are given by 
\begin{align}
    &h_c \coloneqq e^{-\psi_c d_c},\, 
    \label{hc}\\
    &h_a \coloneqq  e^{-\psi_a d_a},
    \label{ha}
\end{align}
where $\psi_c$ and $\psi_a$ are the extinction coefficients that determine the attenuation of the signal (due to scattering or absorption) in the cloud and air, respectively.




\vspace{-4mm}
\subsection{Scintillation Due to Atmospheric Turbulence}

The exponentiated Weibull  distribution has been shown to provide an excellent fit to experimental results for both weak and moderate turbulence conditions compared to lognormal and Gamma-Gamma distributions \cite{Barrios:12}. An attractive feature of exponentiated Weibull distribution is the mathematically tractable nature of its probability density function, which is given by the expression
\begin{small}
\begin{align}
f_{\h_s}(h) = \frac{ab}{\eta} \left( \frac{h}{\eta} \right)^{b-1} \!\!\! \exp \!\left( - \left(\frac{h}{\eta} \right)^b \right) \! \left( 1 - \exp \! \left( \!-\! \left(\frac{h}{\eta} \right)^b \right) \right)^{a-1}\!\!\!, h > 0. \label{ew}
\end{align}
\end{small}
In \eqref{ew}, $b$ is the shape parameter which is representative of the scintillation index. The factor $a$ is also a shape parameter that depends on receiver aperture size, and $\eta$ is the scale parameter. 
\vspace{-8mm}
\subsection{Overall Channel Model}
The overall channel gain---based on channel impairments in \eqref{hrd}, \eqref{hc}, \eqref{ha} and \eqref{ew}---is the product of individual channel impairments: $\h = \h_p \h_s h_a h_c$. Then,  the overall channel probability density function is
{\begin{align}
    f_{\h}(h) = \int_0^{h_ah_cB} \frac{1}{h_ah_cx} f_{\h_p}\left(\frac{x}{h_ah_c} \right)  f_{\h_s}\left( \frac{h}{x} \right)\,dx, \quad h>0.
\end{align}}

\section{Two-Dimensional Position Optimization of UAV for Sensor-to-UAV Link} \label{2-D optimization}
We first introduce the UAV location optimization as a two-dimensional problem in which we optimize the $(x, y)$ coordinates of UAV location to maximize the sum rate from the ground sensors. We assume that the height $z$ of the UAV from the ground is fixed. This implies that  $(x, y, z)$ is the three dimensional position of the UAV. Two dimensional location optimization holds for scenarios when the UAV does not need solar energy to sustain itself in the air, and the UAV is suspended at the smallest possible height from the ground to establish a clear line-of-sight (LoS) with the OGS. The results obtained in this section will be utilized in later sections where we optimize the position of an energy constrained, solar-powered UAV relay.


Let us assume that the uplink radio channel between each sensor and the UAV is dominated by the LoS component. Under the Gaussian channel assumption, the maximum achievable rate (in nats/s/Hz) in the $m$th uplink channel---between the $m$th sensor and the UAV---is $
    C_m = \ln \left( 1 + \frac{P_m G_m}{\sigma_0^2}  \right),$
where $P_m$ is the power transmitted by the $m$th sensor, and $G_m$ is the power gain of the channel for the case when the line-of-sight dominates other paths between the sensor and the UAV \cite{Diamantoulakis:18}. The parameter $\sigma_0^2$ represents the thermal noise power at the UAV antenna. The quantity $G_m$ is defined as 
\begin{align}
    G_m \coloneqq \frac{\beta_0}{z^2 + { (x_m-x)^2 + (y_m-y)^2}},
\end{align}
where $z$ is the height of the UAV from ground in meters, $(x_m, y_m)$ is the location of the $m$th sensor on ground and $\sqrt{ (x_m-x)^2 + (y_m-y)^2}$ is the distance of the $m$th sensor from the UAV. The quantity $\beta_0$ represents the channel power gain at the reference distance of 1 meters from the sensor. The sum rate in the aggregate sensor channel, denoted by $C_s$, is
    $C_s \coloneqq \sum_{m=0}^{M-1} C_m,$
where $M$ is the total number of sensors deployed in the region. Let us assume that the $m$th sensor is located at position $x_m$ for $0 \leq m \leq M-1$. We then have that 
\begin{align}
    C_s = \sum_{m=0}^{M-1}  \ln \left( 1 + \frac{P_m \beta_0 }{\sigma_0^2 \left[z^2 + (x-x_m)^2 + (y-y_m)^2 \right]}  \right).\label{cf4}
\end{align}

\vspace{-8mm}
\subsection{Optimization Problem 1} \label{opt_prob_1}
\vspace{-4mm}
\begin{equation}
\begin{aligned}
 \underset{x, y }{\text{maximize }} 
& \quad  C_s   \\ 
 \text{subject to } & \quad x > 0, y > 0.  \\
\end{aligned} 
\label{opimization1}
\end{equation}

\begin{figure*}
    \centering
    \begin{minipage}[t]
    {\linewidth}
        \centering
        \vspace{-4mm}
        \includegraphics[width = 5.6in]{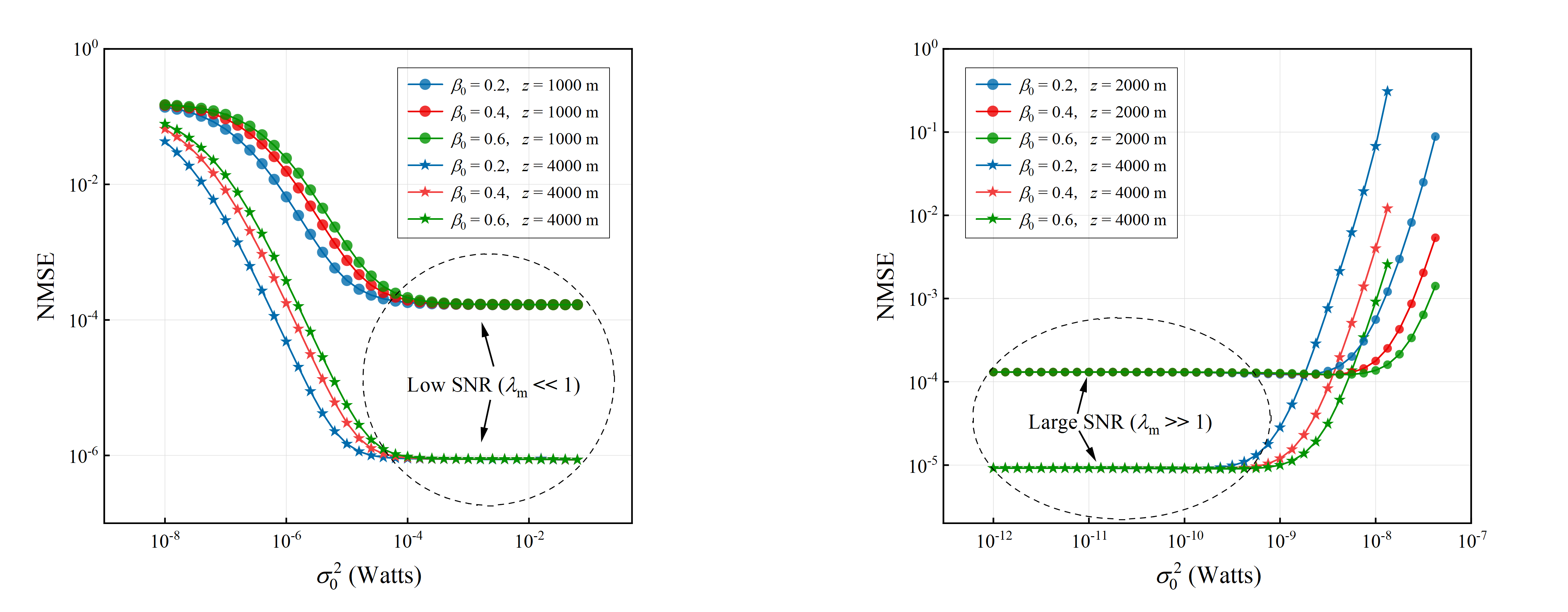}
        \caption{This figure shows the normalized mean-square error as a function of $\sigma_0^2$ for low (left) and large (right) SNR scenarios with different values of factor $\beta_0$ and the height of UAVs $z$. The total link distance $D$ = $1000$ m, the number of the sensors is $M=10$ and the height of UAV $z = 5000$ m.}
        \label{fig:op1_appro}
    \end{minipage}
\end{figure*}


\subsubsection{SNR reasonably large}\label{final 1}
Let us assume that $P_m$ is large enough so that the SNR in the $m$th sensor channel, denoted by $\lambda_m$, is
\begin{align}
\lambda_m \coloneqq \frac{P_m \beta_0 }{\sigma_0^2 \left[z^2 + (x-x_m)^2 + (y-y_m)^2 \right]} > 1 \label{assump_1}
\end{align}
for all $m$. 
  The approximation
\begin{align}
    \ln(1+\lambda_m) \approx \gamma \left( (1+\lambda_m)^{\frac{1}{\gamma}} - 1 \right), \label{Ap1}
\end{align}
holds for a large enough number $\gamma$ \cite{Bashir:TWC:22}. Equation~\eqref{Ap1} can be rewritten as
$
    \ln(1+\lambda_m) \approx \gamma \left( \lambda_m^{\frac{1}{\gamma}} \left( 1 + \frac{1}{\lambda_m} \right)^{\frac{1}{\gamma}} -1 \right).
$
Since $\frac{1}{\lambda_m} < 1$ and $\frac{1}{\gamma} \frac{1}{\lambda_m} \ll 1$, we can use binomial theorem to approximate $\left( 1 + \frac{1}{\lambda_m} \right)^{\frac{1}{\gamma}} \approx 1 + \frac{1}{\gamma \lambda_m}$. Thus,
\begin{align}
    \ln(1+\lambda_m) \approx \gamma \left( \lambda_m^{\frac{1}{\gamma}} \left( 1 + \frac{1}{\gamma \lambda_m} \right) - 1 \right) = \gamma \lambda_m^{\frac{1}{\gamma}} + \lambda_m^{\frac{1}{\gamma} - 1} - \gamma. \label{approx_1}
\end{align}
Substituting approximation \eqref{approx_1} into \eqref{cf4}, we have that the final expression for $C_s$ (for reasonably large SNR) is
\begin{align}
    C_s \approx \sum_{m=0}^{M-1} \left(\gamma \lambda_m^{\frac{1}{\gamma}} + \lambda_m^{\frac{1}{\gamma} - 1} - \gamma \right). \label{C_s_1}
\end{align}
Now, taking the partial derivatives of $C_s$ (Equation~\eqref{C_s_1}) with respect to $x$ and $y$ and setting the derivatives equal to 0, we have that

\begin{align}
    \frac{\partial}{\partial x} C_s& =  \sum_{m=0}^{M-1}  \frac{\sigma_0^2}{P_m \beta_0} (x - x_m) \nonumber \\
    &- \sum_{m=0}^{M-1} \frac{1}{\left( z^2 + (x - x_m)^2 + (y-y_m)^2 \right)} (x - x_m) = 0,  \label{opt10} \\
    \frac{\partial}{\partial y} C_s& =  \sum_{m=0}^{M-1}  \frac{\sigma_0^2}{P_m \beta_0} (y - y_m) \nonumber \\
    &- \sum_{m=0}^{M-1} \frac{1}{\left( z^2 + (x - x_m)^2 + (y-y_m)^2 \right)} (y - y_m) = 0. \label{opt10y} 
\end{align}

When $z$ is large enough so that $z^2 \gg (x - x_m)^2+(y - y_m)^2$ for all $m$, then a solution to \eqref{opt10} and \eqref{opt10y} is obtained as
\begin{align}
    x^* &= \frac{\sum_{m=0}^{M-1} x_m \left( \frac{\sigma_0^2}{P_m \beta_0} - \frac{1}{z^2} \right)}{\sum_{m=0}^{M-1}  \left( \frac{\sigma_0^2}{P_m \beta_0} - \frac{1}{z^2} \right)}, \label{approx1} \\
    y^* &= \frac{\sum_{m=0}^{M-1} y_m \left( \frac{\sigma_0^2}{P_m \beta_0} - \frac{1}{z^2} \right)}{\sum_{m=0}^{M-1}  \left( \frac{\sigma_0^2}{P_m \beta_0} - \frac{1}{z^2} \right)}. \label{approx2}
\end{align}

\subsubsection{Low SNR scenario}\label{final 2}
When $\lambda_m \ll 1$, we have that $\ln \left( 1 + \lambda_m \right) \approx \lambda_m$ and $C_s \approx \sum_{m=0}^{M-1} \lambda_m$. In this case, it can be shown that for $z^2 \gg (x-x_m)^2$ for all $m$, the optimal value of $x$ is
\begin{align}
    x^* &\approx \frac{\sum_{m=0}^{M-1}   P_m{x_m} }{\sum_{m=0}^{M-1}   P_m  }, \label{approx3}\\
    y^* &\approx \frac{\sum_{m=0}^{M-1}   P_m{y_m} }{\sum_{m=0}^{M-1}   P_m  }. \label{approx4}
\end{align}

\subsubsection{Approximation Error Plots} 
Here, we consider the error in  approximate solutions \eqref{approx1} \eqref{approx2}, \eqref{approx3} and \eqref{approx4} to the optimization problem \eqref{opimization1}. Let the exact (numerical) solution to \eqref{opimization1} be denoted by $(\tilde{x}, \tilde{y})$. We then use the normalized mean-square error to quantify the approximation error:
\begin{align}
    \text{NMSE} = \frac{(x^*-\widetilde{x})^2 + (y^*-\widetilde{y})^2}{\widetilde{x}^2+\widetilde{y}^2},
    \label{NMSE}
\end{align}
where $(x^*, y^*)$ are the analytical (approximate) solutions furnished by \eqref{approx1} and \eqref{approx2} for large SNR; and \eqref{approx3} and \eqref{approx4} for low SNR. 
 Fig.~\ref{fig:op1_appro} illustrates the NMSE for low SNR (left figure) and high SNR (right figure) approximations for varying channel power gains at the reference distance $\beta_0$ and at different UAV altitudes represented by $z$. It is noteworthy that the approximate expressions maintain a high level of  accuracy both at low ($\lambda_m \ll 1$) and high SNR regimes ($\lambda_m \gg 1$), particularly within the regions demarcated by dashed regions. However, for the low SNR scenario, the approximation tends to diminish in accuracy as the SNR elevates, especially for $\sigma_0^2 < 10^{-5}$. Similarly, for the high SNR scenario, the approximation error begins to grow for $\sigma_0^2 > 10^{-9}$. We also note that the variation in $\beta_0$ seems to have a nominal effect on the NMSE for both SNR extremities. Yet, a declining UAV altitude begins to erode the approximations' precision since the approximate solutions were computed based on the assumption $z^2 \gg (x-x_m)^2+(y - y_m)^2$. For the said figure, the location of sensors are $\bm{x_m} = \begin{bmatrix}
        0, 100, 150, 200, 250, 350, 450, 550, 750, 1000
    \end{bmatrix}$, $\bm{y_m} = \begin{bmatrix}
        0, 150, 250, 300, 450, 550, 650, 700, 800, 1000
    \end{bmatrix}$, the power transmitted by the sensor is $\bm{P_m} = \begin{bmatrix}
        0.5, 1.0, 1.5, 2.0, 2.5, 3.0, 3.5, 4.0, 4.5, 5.0
    \end{bmatrix}$.

 Fig.~3 presents  the aggregate sensor channel capacity $C_s$ contingent upon the UAV's spatial positioning.  This emphasizes the significance of strategic UAV placement to enhance channel performance. For this figure, the default system parameters are the same as Fig.~\ref{fig:op1_appro}.
 

\begin{figure}
    \centering    \includegraphics[width = 0.75\columnwidth]{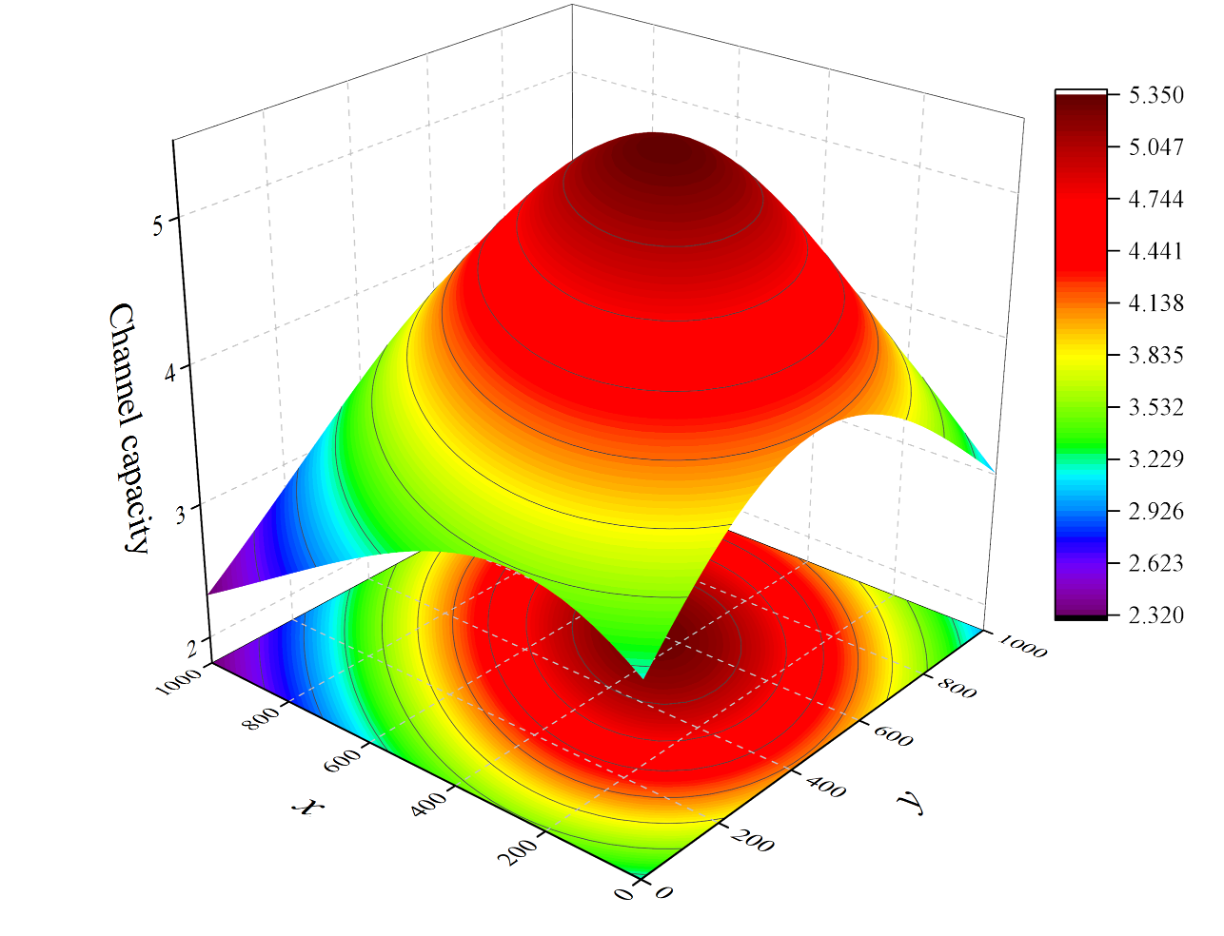}
    \vspace{-4mm}
    \caption{The figure shows 3-D plot of the aggregate sensor-to-UAV channel capacity as a function of UAV's location in two dimensions. The UAV height $z = 500$ m and $\beta_0 = 0.4$. The noise power $\sigma_0^2 = 1\times 10^{-6}$ W.}
    \label{fig}
\end{figure}

\vspace{-2mm}
\subsection{Optimization Problem 2} \label{opt_prob_2}

\begin{equation}
\begin{aligned}
 \underset{x, y }{\text{maximize }}
& \quad   \min_m C_m   \\
 \text{subject to } & \quad x \geq x_0, \, y \geq  y_0. \label{max_min_2D} \\
\end{aligned} 
\end{equation}

A closed-form solution results for the above mentioned problem results when the sensors are arranged in a nondecreasing power sequence from left to right. This arrangement implies that $P_0 \leq P_1 \leq \dotsm \leq P_{M-1}$. This is a natural arrangement in that we place the lower power sensors closer to the OGS and the larger power sensors further away. For a better understanding of problem \eqref{max_min_2D}, we will first consider the solution to one-dimensional problem in terms of $x$ (by fixing the value of $y$) and then extend the solution to the two-dimensional problem for $(x, y)$. The one-dimensional analog of \eqref{max_min_2D} is  
\begin{equation}
\begin{aligned}
 \underset{x }{\text{maximize }}
& \quad   \min_m C_m   \\
 \text{subject to } & \quad x \geq x_0. \label{max_min_1D} \\
\end{aligned} 
\end{equation}

\subsubsection{{Solution of One-Dimensional Problem}}
For the one-dimensional optimization problem, we fix the value of $y$ and maximize capacity as a function of only $x$ coordinates. For the sake of brevity, we represent the uplink capacity of the $m$th sensor, $C_m(x, y)$, as $C_m(x)$ to highlight the one-dimensional nature of the problem. 

For the one-dimensional problem, the solution is given by the following two-step algorithm:
\begin{enumerate}
    \item For a fixed value of $y$, find the set $\mathcal{X}$ of non-negative UAV positions: $\mathcal{X} \coloneqq  \{ x^{(1)}, x^{(2)}, \dotsc, x^{(M-1)} \}$,  such that 
    \begin{align}
        C_0(x^{(i)}) = C_i(x^{(i)}), 
    \label{maxmin}
    \end{align}
    for $i = 1, 2, \dotsc, M-1.$ 
    \item The solution of optimization problem \eqref{max_min_1D} is $x^* = \max \mathcal{X}$. The maximized value of minimum capacity is $C^* = C_0(x^*)$.
\end{enumerate}

\paragraph{Proof}
For the nondecreasing power arrangement, let us define the sequence of $M$ points $x^{(i)}>0$, for $i=0, 1, 2, \dotsc, M-1$. The point $x^{(i)}$ is defined as the position of the UAV relay where the capacity $C_i$ is equal to $C_0$ (or the point where the two capacity curves, $C_i$ and $C_0$, intersect):
\begin{small}
\begin{align}
&C_i\left(x^{(i)}\right) = C_0\left(x^{(i)}\right)  \implies \frac{P_i}{z^2 + \left(x^{(i)}-x_i \right)^2} = \frac{P_0}{z^2 + \left(x^{(i)}-x_0\right)^2} \nonumber \\
& \implies \left(x^{(i)}\right)^2 \left( P_i - P_0 \right) + x^{(i)} \left( 2 P_0 x_i - 2 P_i x_0 \right) \nonumber \\
& \quad \quad \quad \quad \quad \quad \quad \quad+ P_ix_0^2 - P_0x_i^2 + \left( P_i - P_0 \right) z^2 = 0. \label{quad}
\end{align}
\end{small}
The equation~\eqref{quad} is a quadratic in $x^{(i)}$ and can be solved by using the well-known quadratic formula:
$
    x^{(i)} = \frac{-b + \sqrt{b^2-4ac}}{2a},
$
where $a \coloneqq P_i - P_0$, $b\coloneqq 2 \left(P_0 x_i -  P_i x_0 \right)$ and $c \coloneqq P_ix_0^2 - P_0x_i^2 + \left( P_i - P_0 \right) z^2$.

Before we move further, we need to address an important point. Since the curve $C_0$ corresponds to the smallest power, it will determine the minimum rate beyond a certain point $\hat{x}$, where $\hat{x} > x_{M-1}$. For some $x > \hat{x}$, $ \underset{m}{\min} \, C_m(x) =C_0(x)$ since $P_0 < P_m$  and $|x_0 -x| > |x_m -x|$ for any $m > 0$. In this case, $C_0(x)$ is a monotonically decreasing function in $x$ for $x > \hat{x}$.

However, if $x$ approaches $-\infty$ from the point $\hat{x}$, $C_0(x)$---and therefore $\min (C_m)$---will increase and intersect some $C_j(x)$, $j \neq 0$ at point $x = x_j$\footnote{It is possible that $C_0(x)$ may intersect more than one curve at the point $x_j$. However, this does not affect the final result of our argument.}. The curve $C_j(x)$ is monotonically decreasing for $x < x_j$ because if it were increasing for $x < x_j$, the intersection would not have happened in the first place. Thus, for $x < x_j$, $\underset{m}{\min} \, C_m(x) = C_j(x) < C_0(x_j)$. Thus, $\underset{m}{\min}\, C_m$ achieves its maximum at the point $x_j$ and $x^* = x_j = \underset{i}{\max}\,  x^{(i)}, \quad 1 \leq i, j \leq M-1$. It follows that the (maximized) minimum capacity at the optimal point $x^*$ is $ C_0(x^*)$ or $C_j(x^*)$.

\paragraph{Special Case} \label{hahhaa}If $C_0 = \underset{i}{\min}\, C_i$ at $x = x_0$, then no intersection of $C_0$ will take place with any curve  $C_j$, $j = 1, \dotsc, M-1$ for any $x > x_0$. In this case, the optimal UAV location is $x^* = x_0$. 

\subsubsection{Solution of Two-Dimensional Problem}
The one-dimensional optimization problem can be extended in a straightforward manner to the two-dimensional setting. In this regard, we define the distance $D_i$ as the Euclidean distance between the the $0$th and the $i$th sensors. Furthermore, let us define the vector $\bm{d}_i$ as the (variable) vector that points from sensor~$0$ to sensor~$i$. Let us define $d_i \coloneqq \| \bm{d}_i \|_2$ where $0 < d_i \leq D_i$. Now, let the point of intersection between $C_0$ and $C_i$ be given by 
\begin{small}
\begin{align}
     &C_0(d_i) = C_i(D_i-d_i) \implies \frac{P_0}{z^2 + d_i^2} = \frac{P_i}{z^2+(D_i-d_i)^2} \\
     & \implies (P_0 - P_i)d_i^2 - 2P_0 D_i d_i + P_0z^2 - P_iz^2 + P_0D_i^2 = 0, \\
     & \implies d_i^* = \frac{-b - \sqrt{b^2-4ac}}{2a},
\end{align}
\end{small}
where $a \coloneqq P_0 - P_i, b \coloneqq -2P_0D_i$ and $c \coloneqq P_0z^2 - P_iz^2 + P_0D_i^2$. The solution  $\frac{-b + \sqrt{b^2-4ac}}{2a}$ is discarded since it is negative. Then, the two-dimensional point,  where $C_0$ and $C_i$ intersect,  is 
\begin{align}
    \begin{bmatrix}
    x^{(i)} \\ y^{(i)}
    \end{bmatrix} = \begin{bmatrix}
    x_0 \\ y_0
    \end{bmatrix} + \begin{bmatrix}
    \delta x_i^* \\ \delta y_i^*
    \end{bmatrix},
\end{align}
where $\delta x_i^* \coloneqq d_i^* \cos(\theta_i)$ and $\delta y_i^* \coloneqq d_i^* \sin (\theta_i)$ where $\theta_i$ is the angle that $\bm{d}_i$ makes with $x$-axis.
Here, the optimal solution (optimal location of the UAV) for the three dimensional max-min problem is $(x^{(j)}, y^{(j)})$ where $ j = \underset{i}{\argmax} \| \bm{x}^{(i)} \|_2$, and $\bm{x}^{(i)}$ is defined as $\bm{x}^{(i)} \coloneqq \begin{bmatrix}
x^{(i)} & y^{(i)}
\end{bmatrix}^T$. 

\subsubsection{Optimization plots of \eqref{max_min_1D}}

\begin{figure*}
    \centering
    \begin{minipage}[t]
    {\linewidth}
        \centering
        \vspace{-9mm}
        \includegraphics[width = 5.6in]{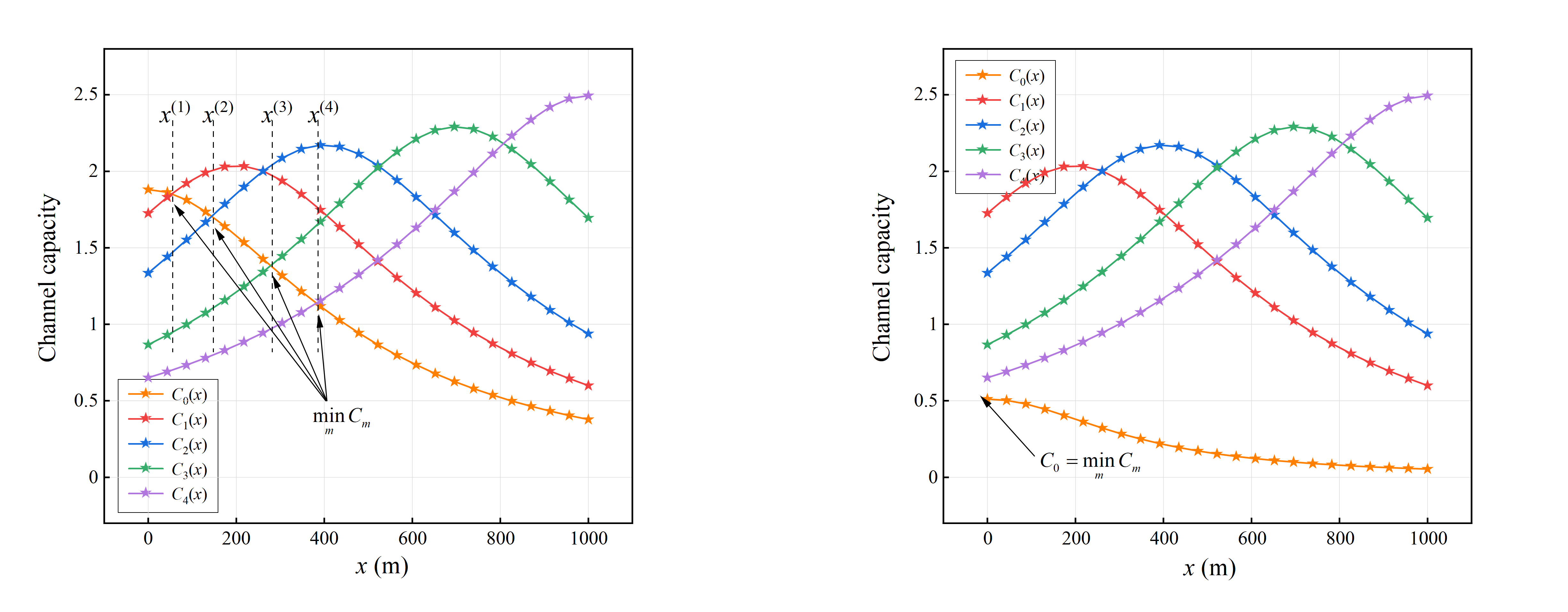}
        \caption{This figure shows the Capacity Curves for Max-min fairness problem.}
        \label{OP222}
    \end{minipage}
\end{figure*}

Fig. \ref{OP222} illustrates the maximum achievable rate (in nats/s/Hz) in the uplink channel between the $m$th sensor and the UAV, denoted as $C_m$, for five different sensors. This problem is formulated as a Max-min optimization problem. From the graph, $C_0$ curve---represented by orange color---intersects with other four curves at points $x^{(i)}$, where $i = 1,2,3,4$. Among these intersections, the point $x^{(4)}$ notably produces the largest minimum value for $C_m$ across all sensors. For Fig.~\ref{OP222}, the parameter values are as follows: The total link distance $D$ = $1000$ m, the number of the sensors is $M=5$, the location of sensors are $\bm{x_m} = \begin{bmatrix}
        0, 200, 400, 700, 1000
    \end{bmatrix}$, the power transmitted by the sensors is $\bm{P_m} = \begin{bmatrix}
        2.5, 3, 3.5, 4, 5
    \end{bmatrix}$ (left) and $\bm{P_m} = \begin{bmatrix}
        0.3, 3, 3.5, 4, 5
    \end{bmatrix}$ (right), the factor $\beta_0 = 0.2$, the noise power $\sigma_0^2 = 1\times 10^{-6}$ and the height of UAV $z = 300$ m.

As for the right graph in Fig. \ref{OP222} represents to the special case previously discussed in \ref{hahhaa}. Here, $C_0$ corresponds to  $ \displaystyle \min_m C_m$ for all $x > x_0$. Here, $C_0$ does not intersect with any of the four remaining curves. As a result, we find that the optimal point that maximizes $\min C_m$ coincides with $x_0$.


\section{Solar Energy Harvesting and Energy Consumption Models of Hovering UAVs}\label{energy harvesting models}
\subsection{Energy Harvesting Model}
In this paper, we assume that the UAV is fitted with a large enough solar panel so that the energy required by the UAV rotors as well as laser transmission to the OGS is provided solely by  sunlight. In this case, the atmospheric transmittance at an altitude $z$ in the region above the clouds is determined as \cite{sun2019optimal}
\begin{align}
    \alpha_0(z) = A_0 - B_0 \exp\left(-\frac{z}{\delta} \right),
\end{align}
where $A_0$ is the maximum value of atmospheric transmittance, $B_0$ is the extinction coefficient of the atmosphere in m$^{-1}$, and $\delta$ is the scale height of the earth in m. Besides this, the presence of clouds and the dirt in air will further reduce the sunlight energy reaching the UAV. In this regard, let us define the energy attenuation through clouds and dirt as
    $\alpha_c(Z_c)  \coloneqq \exp\left( - \beta_c Z_c \right)$ and $\alpha_d(Z_d)  \coloneqq \exp\left( - \beta_a Z_d \right)$. The quantities $\beta_c$ and $\beta_a$ denote the extinction coefficient of the cloud and the air mediums, respectively. Then,  the harvested solar power at the UAV, denoted by $P_H$, is given by \cite{sun2019optimal}
\begin{align}
    P_H(z) = \begin{cases}
    \eta_p A G_r \alpha_0(z), & z > Z_c + Z_d, \\
    \eta_p A G_r \alpha_0(z) \alpha_c(Z_c + Z_d - z), & Z_d <z < Z_c + Z_d, \\
    \eta_p A G_r \alpha_0(z) \alpha_c(Z_c) \alpha_d(Z_d-z), & z <  Z_d, \\
    \end{cases}
    \label{PHH}
\end{align}
where $\eta_p$ is the photoconversion efficiency of photovoltaic cells, $A$ is the active surface area of the photovoltaic array in m$^2$ and $G_r$ denotes the average solar radiation on earth in W/m$^2$.

\vspace{-2mm}
\subsection{Energy Consumption Model}

It has been shown in \cite{Abeywickrama:18:Access} that the power consumed by a UAV  while hovering is
\begin{align}
    P_0 = \sqrt{\frac{(mg)^3}{2 \pi r^2 \rho}},
\end{align}
where $m$ is the mass of the UAV and the solar panels in kg, $g$ is the gravitational acceleration: $g \coloneqq 9.8$ m/s, $r$ is the radius of the propeller in meters, and $\rho$ is the density of air at a given height and temperature in kg/m$^3$. For instance, for a UAV with a total mass of 5 kg, propeller radius of 0.5 m and the air density $\rho = 1.225$ kg/m$^3$, the power required to sustain the UAV in the hovering position is approximately 247 Watts.  Thus, the height of the UAV ($z$) and the area of the solar panel, $A$, has to be such that the harvested power $P_H(z)$ has to be more than 247 Watts for this particular scenario.

With the advancements in photovoltaic technology, the power-to-weight and the power-to-area ratio of the solar arrays has improved considerably over the last decade. For instance, the flexible textile solar panels devised by SolarCloth System (France) can deliver power densities up to 170 W/m$^2$ with an average weight density of 500 g/m$^2$ \cite{Power-Weight-Ratio-Solar-Panels}. This leads to an energy density of 340 W/kg with the textile solar panels. The United States based Ascent Solar Technologies developed super lightweight high-quality crystalline silicon solar arrays in 2016 that can deliver energy density up to 1700 W/kg. The highly flexible perovskite solar arrays have been shown to achieve a power density of 23 W/g \cite{Kaltenbrunner:15:Nature}. These high power-to-weight and power-to-area ratio arrays are highly suitable for deployment on unmanned aerial vehicles, weather balloons and aeroplanes \cite{Kaltenbrunner:15:Nature}.


\section{End-to-End Rate of the Sensor-UAV-OGS Link for Solar UAV} \label{end-to-end optimization}
As discussed before, the UAV assumes the role of a relay to connect the sensor field with the OGS: The sensor-to-UAV link is RF whereas the UAV-to-OGS link is optical wireless. In this section, we consider the end-to-end channel capacity for this hybrid RF/FSO link that utilizes a solar energy harvesting amplify-and-forward (AF) or decode-and-forward (DF)  UAV relay. We denote the capacity in the laser backhaul (UAV-to-OGS link) by the symbol $C_b$.
\vspace{-2mm}
\subsection{AF Relay}\label{af_relay}
Let $X$ denote the sum signal reaching the AF relay from the sensor network. We have that
\begin{align}
    X = \underbrace{\sum_{m=0}^{M-1}  \sqrt{\frac{P_m \beta_0}{z^2 + (x - x_m)^2+(y-y_m)^2}}}_{S_X} + \underbrace{\sum_{m=0}^{M-1} N_m}_{N_X}, \label{signal_X}
\end{align}
where $N_m$ are the i.i.d. Gaussian random variables: $N_m \sim \mathcal{N}\left(0, \sigma_0^2 \right)$ where $\sigma_0^2$ is the thermal noise variance in a single sensor channel. Let $\sigma_s^2$ denote the total noise variance in the aggregate sensor channel. Then,  $\sigma_s^2 = M \sigma_0^2$ where $M$ is the total number of sensors. By definition, $N_X \sim \mathcal{N}(0, \sigma_s^2)$.

The laser signal retransmitted by the AF relay to the OGS is $Y = G X$, where $G$ is the voltage amplification factor. Let us assume that the average harvested power is $P_H$ Watts and the average power to sustain the UAV in the air is $P_0$ Watts. Then, the net power available for the purpose of signal transmission is $P_H- P_0$. Since the relay cannot transmit more than the net power,  we have that 
\begin{small}
\begin{align}
    P_H-P_0 & \geq  \E [G^2 X^2] = G^2 \E[X^2] \nonumber \\
    &= G^2\E[S_X^2 + N_X^2 + 2 S_X N_X] = G^2S_X^2 + G^2\sigma_s^2 \\
    & \implies G^2 \leq \frac{P_H - P_0}{S_X^2 + \sigma_s^2} \implies G \leq \sqrt{\frac{P_H - P_0}{S_X^2 + \sigma_s^2}}.
\end{align}
\end{small}
Here, for the purpose of analysis, we set $G$ equal to the maximum possible value it can attain: i.e., $G = \sqrt{\frac{P_H-P_0}{S_X^2 + \sigma_s^2}}.$

The received signal at the OGS is denoted by $Z$. We have that
\begin{small}
\begin{align}
    Z &= \eta \mathbbm{h} Y + N = \eta \mathbbm{h} G X + N = \eta \mathbbm{h} G (S_X + N_X) + N \nonumber\\
    & = \eta \mathbbm{h} G S_X + \eta \mathbbm{h} G N_X + N.
\end{align}
\end{small}
where $\mathbbm{h}$ is the (random) optical channel gain in the backhaul, $\eta$ is the photoconversion efficiency at the OGS receiver and $N$ is a Gaussian random variable which represents thermal noise at the OGS optical receiver: $N \sim \mathcal{N}(0, \sigma_N^2)$.
 The conditional end-to-end rate (in bits/seconds/Hz) for the AF relay is
\begin{align}
    C({h}) &= \ln \left( 1 + \frac{(\eta {h} G S_X)^2}{(\eta {h} G)^2 \sigma_s^2 + \sigma_N^2} \right) \nonumber \\
    &= \ln \left( 1 + \frac{\eta^2 h^2 \left(\frac{P_H-P_0}{S_X^2+ \sigma_s^2}\right)  S_X^2}{\eta^2 h^2 \left(\frac{P_H-P_0}{S_X^2+ \sigma_s^2}\right) \sigma_s^2 + \sigma_N^2} \right) \label{S_X1}
\end{align}
 Finally, the unconditional rate (averaged over $\mathbbm{h}$) is $C = \int_0^\infty C(h) f_{\mathbbm{h}}(h) \, dh$.


\subsubsection{Asymptotic Case 1}\label{yabai}
When the SNR in the aggregate sensor-to-UAV channel is high, i.e., $S_X^2 \gg \sigma_s^2$ and the SNR in the UAV-to-OGS optical channel is low, i.e., $ \sigma_N^2 \gg \eta^2 h^2 \left(\frac{P_H-P_0}{S_X^2+ \sigma_s^2}\right) \sigma_s^2$, then, 
\begin{align}
    C(h)& = \ln \left( 1 + \frac{\eta^2 h^2 \left(\frac{P_H-P_0}{S_X^2+ \sigma_s^2}\right)  S_X^2}{\eta^2 h^2 \left(\frac{P_H-P_0}{S_X^2+ \sigma_s^2}\right) \sigma_s^2 + \sigma_N^2} \right) \nonumber \\
    &\approx \ln \left( 1 + \frac{\eta^2 h^2 \left(P_H-P_0\right)}{ \sigma_N^2} \right). \label{asymptot_AF_1}
\end{align}
In this scenario, we note that the backhaul channel dominates the end-to-end capacity. To understand the effect of link distance on the capacity,  we note that for the fading scenario, the signal strength's dependence on link distance is only through pointing error\footnote{It is shown later in the same section that the loss in end-to-end capacity due to pointing error increases directly with link distance.} and not scintillation. Therefore, without loss of generality, we assume that the optical channel suffers fading mainly due to pointing error. Under this assumption, the composite channel gain is $\h = \h_p h_a h_c$ with PDF: 
$f_{\h} (h) = \frac{\Phi}{h_a h_c} \left( \frac{h}{h_a h_c} \right)^{\frac{\theta^2-\sigma^2}{\sigma^2}} \cdot \mathbbm{1}_{[0,Bh_ah_c)}(h).$

In this case, the unconditional rate is 
\begin{small}
\begin{align}
    C &\approx \int_0^{Bh_ah_c} \ln \left( 1 + \frac{\eta^ 2 {h^2} (P_H-P_0)}{\sigma_N^2}  \right) \cdot \frac{\Phi}{h_a h_c} \left( \frac{h}{h_a h_c} \right)^{\frac{\theta^2-\sigma^2}{\sigma^2}}  \, dh  
    \label{C_appro1}
\end{align}
\end{small}
Recall that from \eqref{Phi}, $\Phi = \frac{\theta^2}{\sigma^2} \left( \frac{1}{B} \right)^{\frac{\theta^2}{\sigma^2}}$. To get a closed-form expression, we assume the common result that $\theta^2 \approx 2\sigma^2$, which is justified in \cite{Bashir:TWC:22}. Under this assumption, \eqref{C_appro1} can be further simplified as
\begin{small}
\begin{align}
    C &\approx \int_0^{Bh_ah_c} \frac{2}{B^2h_a^2 h_c^2} \cdot \ln \left( 1 + \frac{\eta^2  (P_H - P_0)h^2}{\sigma_N^2}  \right) \cdot h \, dh \nonumber \\
    &= \left( 1 + \frac{\sigma_N^2}{\eta^2 h_a^2h_c^2 (P_H-P_0)B^2} \right)\nonumber \\
    & \quad \quad \quad \quad \quad\cdot\ln\left( 1 + \frac{\eta^2 h_a^2 h_c^2 B^2 (P_H - P_0)}{\sigma_N^2} \right) - 1
\end{align}
\end{small}
In the presence of fading due to pointing error, the relationship between link distance and attenuation due to pointing error is not clear. In this scenario, to show that the channel capacity of the laser link is maximized when the UAV lies as close as possible to the OGS, we first show that $C$ is monotonically increasing in $t$ where $t \coloneqq \frac{\eta^2 h_a^2 h_c^2 B^2 (P_H - P_0)}{\sigma_N^2} > 0$. Then, $C(t) = \left(1+\frac{1}{t}\right)\ln\left(1+t\right) -1$, and by taking the first derivative, we have that $C'(t) = \frac{t - \ln(1+t)}{t^2} > 0,$
since  $t > \ln(1+t) $ for all $t>0$. Thus, $C$ is monotonically increasing in $t$. This implies that---for a fixed height $z$---we want to maximize only $B, h_a$ and $h_c$ to maximize $C$ since the term $\frac{\eta^2 h_a^2 h_c^2 (P_H - P_0)}{\sigma_N^2}$ does not depend on UAV location in $(x, y)$ plane. The quantities $B, h_a$ and $h_c$ are maximized only at $(0, 0, z)$, which is exactly on top of the OGS location on ground. In other words, for a fixed UAV height $z$, the optimal  UAV position in the $(x, y)$ plane is the same as the position of the OGS. 

\vspace{0mm}
\subsubsection{Asymptotic Case 2} \label{AFasymptotic2}
When the SNR in the optical UAV-to-OGS channel is high compared to the aggregate sensor-to-UAV channel such that $ \sigma_N^2 \ll \eta^2 h^2 \left(\frac{P_H-P_0}{S_X^2+ \sigma_s^2}\right) \sigma_s^2$, then the channel capacity is approximated as
\begin{align}
    &C(h) = \ln \left( 1 + \frac{\eta^2 h^2 \left(\frac{P_H-P_0}{S_X^2+ \sigma_s^2}\right)  S_X^2}{\eta^2 h^2 \left(\frac{P_H-P_0}{S_X^2+ \sigma_s^2}\right) \sigma_s^2 + \sigma_N^2} \right)  \nonumber\\
    &\approx  \ln \left( 1 + \frac{\eta^2 h^2 \left(\frac{P_H-P_0}{S_X^2+ \sigma_s^2}\right)  S_X^2}{\eta^2 h^2 \left(\frac{P_H-P_0}{S_X^2+ \sigma_s^2}\right) \sigma_s^2  } \right)  = \ln \left( 1 + \frac{  S_X^2}{ \sigma_s^2  } \right), \label{AF_asymptot2}
\end{align}
which is simply the capacity of the aggregate sensor-to-UAV channel and its maximum value depends only on the positions and transmitted powers of the sensors (the optimum UAV position for this case was derived in Section~\ref{2-D optimization}).
\vspace{0mm}
\subsubsection{Asymptotic Case 3} \label{AFasymptotic3}
For large noise power $\sigma_N^2$ at the OGS receiver such that, $ \sigma_N^2 \gg \eta^2 h^2 \left(\frac{P_H-P_0}{S_X^2+ \sigma_s^2}\right) \sigma_s^2$, the approximate end-to-end capacity is 
\begin{align}
C(h) \approx \ln \left( 1 + \frac{\eta^2 h^2 \left(\frac{P_H-P_0}{S_X^2+ \sigma_s^2}\right)  S_X^2}{ \sigma_N^2} \right). \label{asymptot_AF_2}
\end{align}
 In this asymptotic case, the value of $\sigma_N^2$ does not play any role in determining the optimal UAV position since the optimal position depends only on the numerator.

\vspace{-2mm}
\subsection{DF Relay} \label{df_relay}
The conditional end-to-end rate for a DF relay scheme is given by
\begin{align}
 C (h, x, y, z) = \min( C_b(h, x, y, z), C_s(x, y, z) ).
\end{align}

The dependence of aggregate sensor channel capacity $C_s$ on UAV location $(x, y, z)$ is evident through \eqref{cf4}. The capacity in the backhaul, $C_b$, depends on the UAV location through  the distribution of channel coefficient $\mathbbm{h}_p$ which is a function of $(x, y, z)$ because of the factor $B$ (as depicted in \eqref{B_1_val}). Additionally, $C_b$ also depends on channel coefficients $h_a$ and $h_c$---which are defined by \eqref{ha} and \eqref{hc}, respectively---which also depend on the UAV position. The average end-to-end rate of the DF link is
\begin{align}
    &C(x, y, z) = \int_0^{B h_a h_c} C(h, x, y, z) f_\h(h)\, dh \nonumber \\ 
    & =  \int_0^{B h_a h_c} \min \Bigg( \ln \left( 1 + \frac{((P_H-P_0) h \eta)^2}{\sigma_N^2} \right), \nonumber\\
    &\sum_{m=0}^{M-1}  \ln \left( 1 + \frac{P_m \beta_0 }{\sigma_0^2 \left[z^2 + (x-x_m)^2 + (y-y_m)^2 \right]}  \right) \Bigg) f_{\mathbbm{h}}(h) \, dh. \label{DF_capacity}
\end{align}
We can obtain a closed-form expression for the approximate channel capacity when the pointing error is the major contributing factor to signal fading \cite{Bashir:TWC:22}. In that case, the approximate closed-form expression of capacity, denoted by $\tilde{C}$, is \cite{Bashir:TWC:22}
\begin{small}
\begin{align}
    &\widetilde{C}(x,y,z) \nonumber\\
    &=  \Phi \frac{(P_H-P_0)^2\eta^2}{\sigma_N^2} \frac{\sigma^2}{\theta^2+2\sigma^2}\left(\min(\hat{h},h^{\ast},Bh_ah_c)\right)^{\frac{\theta^2+2\sigma^2}{\sigma^2}} \nonumber \\
    &+ \Phi \sigma^2 \Bigg(\frac{\left(\min\left(\hat{h},h^{\ast},Bh_ah_c\right)-\min\left(h^{\ast},Bh_ah_c\right)\right)\alpha}{\theta^2} \nonumber \\
    &- \frac{\left(\min\left(\hat{h},h^{\ast},Bh_ah_c\right)\right)^{\frac{2\sigma^2+\alpha\theta^2}{\alpha\sigma^2}}\times \alpha\left(\frac{(P_H-P_0 )\eta}{\sigma_N}\right)^{\frac{2}{\alpha}} }{\alpha\theta^2 + 2\sigma^2} \nonumber \\
    &+ \frac{\left( \min(h^{\ast},Bh_ah_c)\right)^{\frac{2\sigma^2+\alpha\theta^2}{\alpha\sigma^2}}\times \alpha\left(\frac{(P_H-P_0 )\eta}{\sigma_N}\right)^{\frac{2}{\alpha}} }{\alpha\theta^2 + 2\sigma^2} \nonumber \\
    &- \frac{ \min(\hat{h},h^{\ast},Bh_ah_c)^{\frac{2\sigma^2+\alpha\theta^2-2\alpha\sigma^2}{2\sigma^2}}  \times \left(\frac{(P_H-P_0 )\eta}{\sigma_N}\right)^{\frac{2-2\alpha}{\alpha}}}{\alpha\theta^2 + 2\sigma^2-2\alpha\sigma^2} \nonumber \\ +
    &\frac{  \min(h^{\ast},Bh_ah_c)^{\frac{2\sigma^2+\alpha\theta^2-2\alpha\sigma^2}{2\sigma^2}}  \times \left(\frac{(P_H-P_0 )\eta}{\sigma_N}\right)^{\frac{2-2\alpha}{\alpha}}}{\alpha\theta^2 + 2\sigma^2-2\alpha\sigma^2}\Bigg) \nonumber \\
    &+ \sum_{m=0}^{M-1}  \ln \left( 1 + \frac{P_m \beta_0 }{\sigma_0^2 \left[z^2 + (x-x_m)^2 + (y-y_m)^2 \right]}  \right) \nonumber \\
    &\times \frac{\Phi\sigma^2}{\theta^2}\left({(Bh_ah_c)}^{\frac{\theta^2}{\sigma^2}} - \Big(\min(h^{\ast},Bh_ah_c)\Big)^{\frac{\theta^2}{\sigma^2}} \right),
\end{align}
\end{small}
where $ h^* \coloneqq \frac{\sigma_N \sqrt{ \prod_{m=0}^{M-1} \left( 1 + \frac{P_m \beta_0 }{\sigma_0^2 \left[z^2 + (x-x_m)^2 + (y-y_m)^2 \right]}\right) - 1}}{(P_H-P_0)\eta},\,\hat{h} \coloneqq \frac{\sigma_N}{(P_H-P_0)\eta}$ and $\alpha \gg 1$.

\subsubsection{Asymptotic Case 1} \label{DFasymptotic1}
When the SNR in the aggregate sensor-to-UAV channel is high compared to the SNR in the optical backhaul UAV-to-OGS channel---that is $\sigma_N^2$ is large---the capacity in the backhaul $C_b$ dominates the end-to-end rate. In this case,  the average end-to-end rate is approximately
\begin{align}
    C(x,y,z) = \int_0^{B h_a h_c}  \ln \left( 1 + \frac{((P_H-P_0) h \eta)^2}{\sigma_N^2} \right) f_{\mathbbm{h}}(h) \, dh, 
\end{align}
which is similar to the asymptotic case 1 for AF scheme in \ref{yabai}. Therefore, for this asymptotic case, the optimal UAV based DF relay position in the $(x,y)$ plane is the same as the position of the OGS.

\subsubsection{Asymptotic Case 2} \label{DFasymptotic2}

When the SNR in the aggregate sensor-to-UAV channel is low compared to the SNR in the optical UAV-to-OGS channel---that is $\sigma_N^2$ is low---then the capacity in the aggregate sensor channel, $C_s$, is dominant. In this case, the average end-to-end rate in \eqref{DF_capacity} can be simplified to
\begin{align}
    &C(x,y,z)
    = \sum_{m=0}^{M-1} \ln \left( 1 + \frac{P_m \beta_0 }{\sigma_0^2 \left[z^2 + (x-x_m)^2 + (y-y_m)^2 \right]}  \right).
\end{align}

In other words, the optimal location of UAV depends only on the positions and transmitted powers of the sensors which is studied in detail in Section~\ref{2-D optimization}.

\begin{figure*}
	\setlength{\abovecaptionskip}{-5pt}
	\setlength{\belowcaptionskip}{-10pt}
	\centering
	\begin{minipage}[t]{0.33\linewidth}
		\centering
		\includegraphics[width=2.6in]{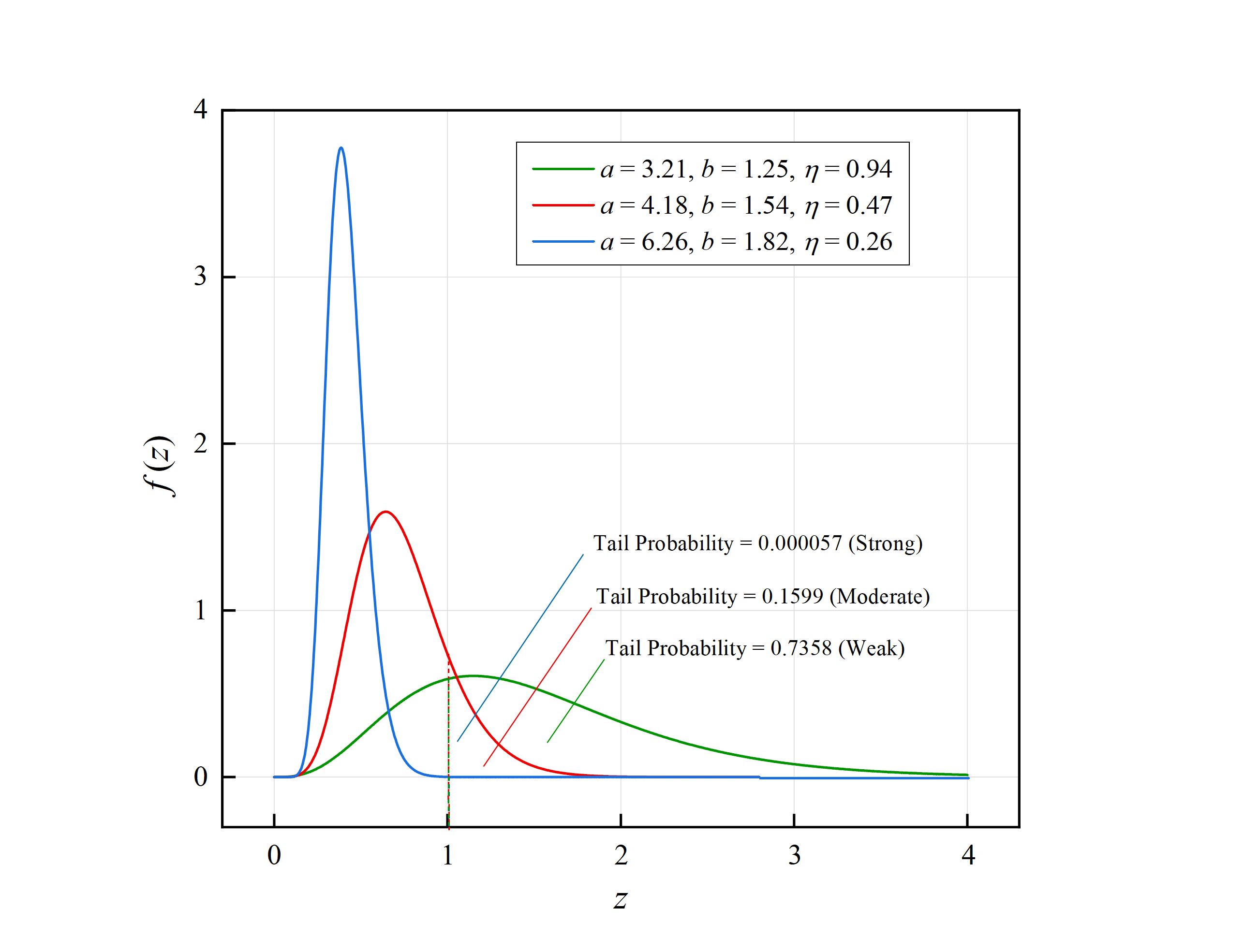}
		\label{figure1}
	\end{minipage}%
	\begin{minipage}[t]{0.33\linewidth}
		\centering
		\includegraphics[width=2.6in]{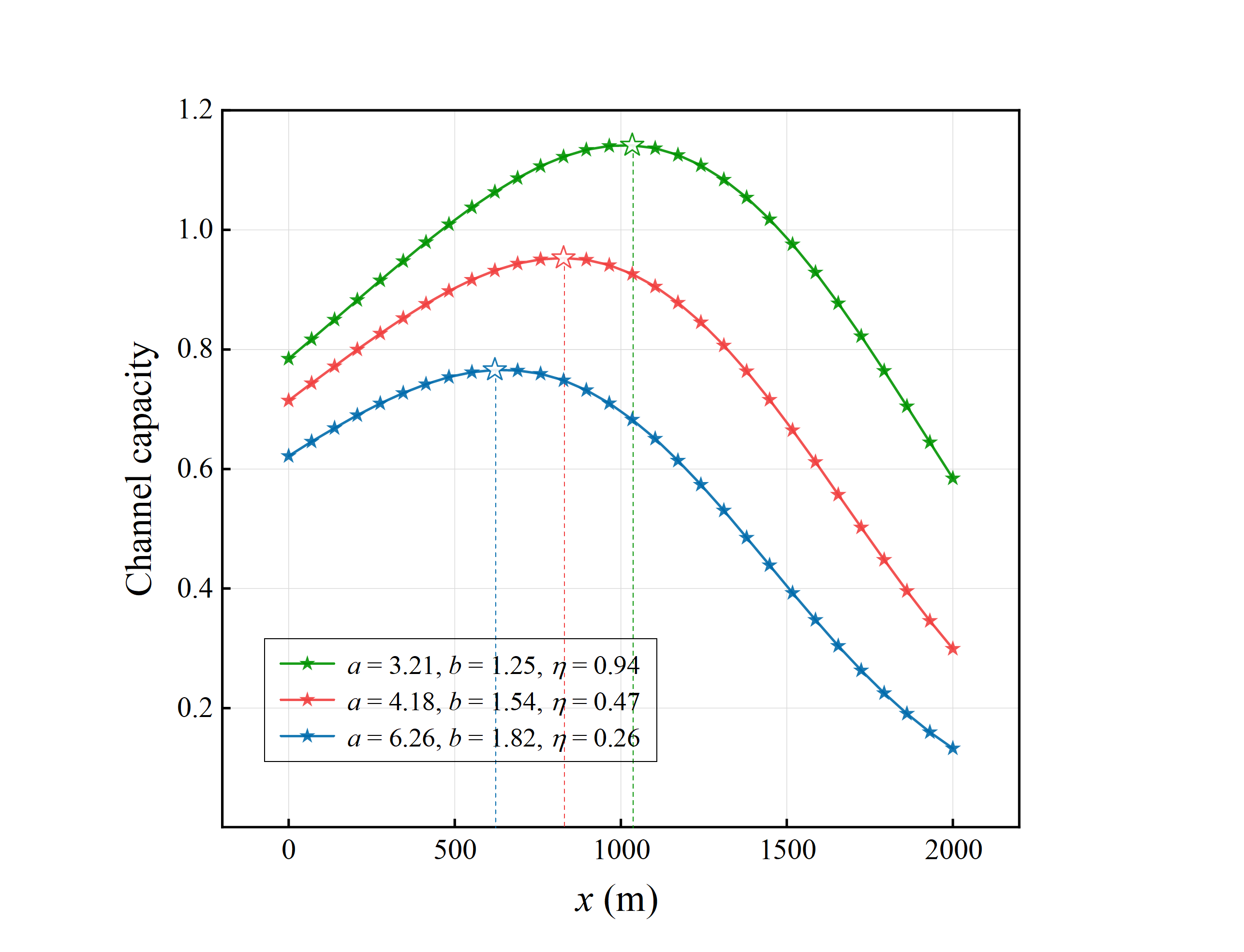}
		\label{figure2}
	\end{minipage}
	\begin{minipage}[t]{0.33\linewidth}
		\centering
		\includegraphics[width=2.6in]{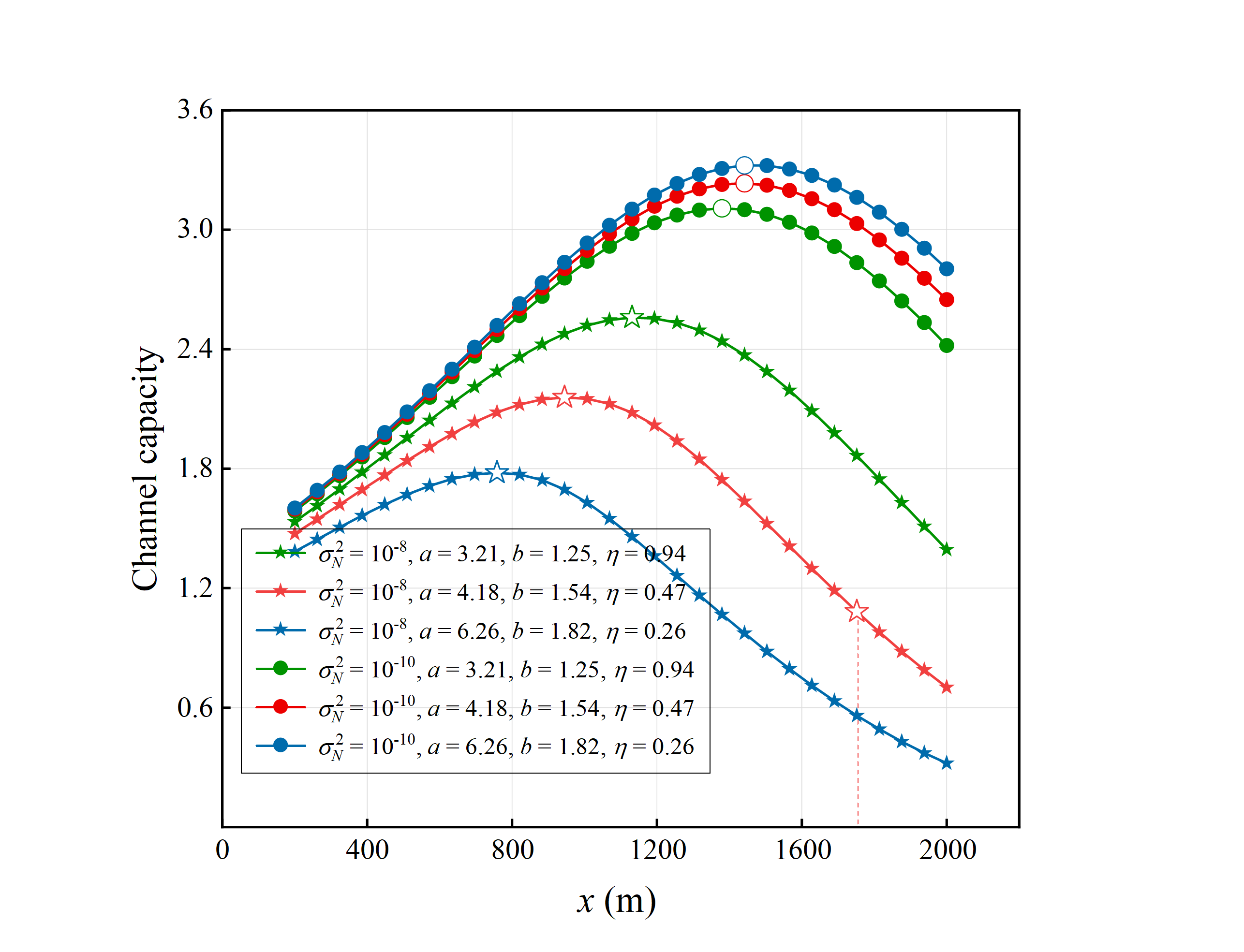}
		\label{figure3}
	\end{minipage}
 \vspace{-4mm}
 \caption{The left figure shows Exponentiated Weibull fading with different parameters. The center and right figures shows capacity plots (as a function of $x$ coordinate of UAV) for AF and DF relays, respectively.}
 \label{OP3_turbulence}
  \vspace{0mm}
        \begin{minipage}[t]
            {\linewidth}
            \centering
            \includegraphics[width=5.6in]{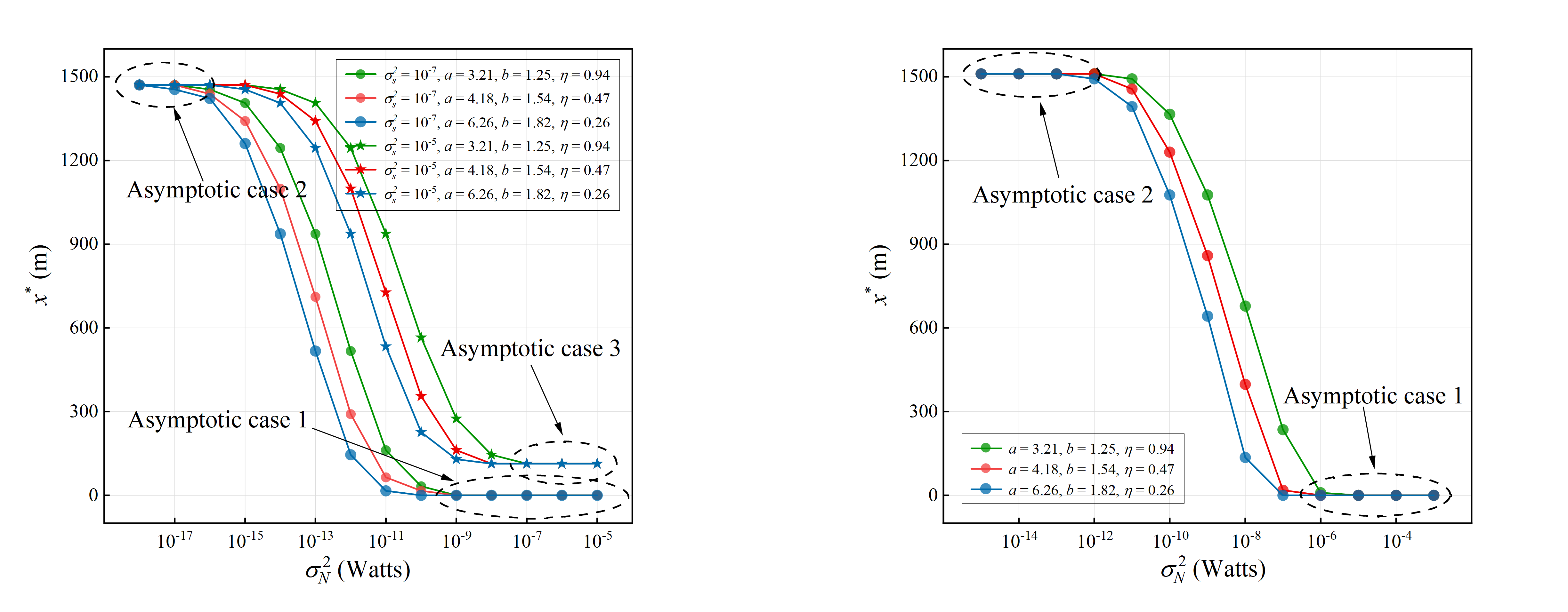}
        \end{minipage}
        \vspace{-4mm}
        \caption{This figure shows optimal location of UAV along $x$ axis as a function of optical noise power $\sigma_N^2$ for AF scheme (left) and DF scheme (right) under  Exponentiated Weibull fading. The attenuation factor $\psi_c = 0.003$.}
        \label{OP3_new3}
\end{figure*}
\section{Three-Dimensional Position Optimization of Solar UAV for Sensor-UAV-OGS Link} \label{3-D optimization}

In our optimization analysis, our primary objective is to enhance the end-to-end  capacity of the sensor-UAV-OGS link by optimizing the solar-powered UAV position in  $x, y$ and $z$ dimensions. The motivation for optimization in $(x, y)$ plane stems from the fact that the end-to-end rate of sensor-UAV-OGS link depends on both the sensor-to-UAV link and the UAV-to-OGS link: If the UAV is too close to the OGS, the sensor-to-UAV link will suffer performance degradation, and if the UAV is too close to the sensor field, the UAV-to-OGS link will underperform. Therefore, an optimal choice of $(x, y)$ coordinates of the UAV is important to maintain an optimum end-to-end rate.

The height $z$ of the UAV is important in that it decides the amount of energy harvested from the sun as well as the performance of sensor-to-UAV and UAV-to-OGS links. If the UAV is placed too high from the ground, the UAV will harvest more solar energy but it will lie further away from the sensors and the OGS which will minimize the end-to-end rate. On the other hand, if the UAV is too close to the ground, the minimized harvested energy will affect the UAV-to-OGS link. Thus, an optimal height $z$ is necessary to guarantee a maximum end-to-end capacity performance. Additionally, recognizing the impact of weather conditions on UAV placement, this study also caters to the affect of clouds on the optimal altitude choice for the UAV.


In this section, we consider the three-dimensional position optimization of a solar-powered UAV relay. For the optimization problem, we have standardized the position of all sensors (in meters) as follows: 
$
\bm{x_m} = \begin{bmatrix}
        700, 800, 900, 1000, 1200, 1400, 1500, 1600, 1800, 2000
    \end{bmatrix} $ and \newline 
    $\bm{y_m} = \begin{bmatrix}
        800, 900, 1000, 1200, 1300, 1500, 1600, 1700, 1900, 2000
    \end{bmatrix}.$
The default parameter values of our optimization problems are enumerated in Table~\ref{table1}. 
\begin{table}

    \centering
    \begin{center}
    \scalebox{0.6}{
    \begin{tabular}{|{c}|{c}|{c}|}
         \hline
         \textbf{Symbol} & \textbf{Parameter} & \textbf{Simulation Value} \\ \hline
         $a_d$ & Radius of OGS receiver telescope (m) & 0.5 \\ \hline
         $\theta$ & Angular beamwidth (radians) & $1\times10^{-2}$ \\ \hline
         $\sigma$ & Angular pointing error standard deviation (radians) & $1\times10^{-2}$ \\ \hline
         $D$ & Distance between OGS and farthest sensor (m) & 1000, 2000 \\ \hline
         $Z_c$ & Distance  sunlight travels through the cloud (m) & 500 \\ \hline
         $Z_d$ & Distance  sunlight travels through  dirt medium (m) & 500 \\ \hline
         $P_m$ & Power transmitted by the $m$th sensor (Watts) & 0.2--5 \\ \hline
         $\beta_0$ & Channel power gain at the reference distance from the sensor & 0.2, 0.4, 0.6 \\ \hline
         $M$ & Total number of sensors & 10 \\ \hline
         $A_0$ & Maximum value of atmospheric transmittance & 0.8978 \\ \hline
         $B_0$ & Extinction coefficient of the atmosphere (m$^{-1}$) & 0.2804 \\ \hline
         $\delta$ & Scale height of the earth (m) & 8000 \\ \hline
         $G_r$ & Average solar radiation on earth (W/m$^{2}$) & 1361\\ \hline
         $A$ & Active surface area of the photovoltaic array (m$^{2}$) & 2 \\ \hline
         $\eta_p$ & Photoconversion efficiency of photovoltaic cells & 0.4 \\ \hline
         $\eta$ & Photoconversion efficiency at the OGS receiver & 0.4 \\ \hline
         {$\psi_a$} & Attenuation/extinction factor of  laser signal  through air & 0.001 \\ \hline
         $\psi_c$ & Attenuation/extinction factor of  laser signal through cloud & 0.001-0.01 \\ \hline
         $\beta_c$ & Attenuation/extinction factor of sunlight through cloud  & 0.001-0.02  \\ \hline
         {$\beta_a$} & Attenuation/extinction factor of sunlight  through air  & 0.5  \\ \hline
         $P_0$ & Power consumed by  hovering UAV (Watts) & 100 \\ \hline
         $\sigma_0^2$ & Noise power of sensor-to-UAV link (Watts) & $1\times10^{-6}$ \\ \hline
         $\sigma_N^2$ & Thermal noise power at the OGS optical receiver (Watts) & $1\times10^{-10}$ \\ \hline
\end{tabular}
}
    \hspace{16mm}
    \caption{\textbf{Table Of simulation parameters}}\label{table1}
    \vspace{-10mm}
    \end{center}    
\end{table}
The UAV position optimization problem is formulated as
\begin{equation}
\begin{aligned}
 \underset{x, y, z }{\text{maximize }} \label{3D_opt}
& \quad  C   \\
 \text{subject to } & \quad i)\, 0 < x < D,\\
 & \quad ii) \, 0 < y < D, \\
 & \quad iii) \, z > z_0.\\
\end{aligned} 
\end{equation}
We employ numerical optimization techniques to solve this optimization problem. Here, the set of sensor's transmission power (in Watts) is given by 
$
\bm{P_m} =[
        0.2, 0.4, 0.6, 0.8, 1, 1.2, 1.4, 1.6, 1.8, 2
    ].$ The constraint $z>z_0$ corresponds to the condition $P_H(z) - P_0 > 0$, where $P_H(z)$ is a function of  the UAV's altitude.

We now discuss the effect of UAV position on the end-to-end capacity through graphical illustrations. For the purpose of clarity, we only consider 2-D plots (functions of only one variable) to convey our results more effectively.

Fig.~\ref{OP3_turbulence} illustrates the maximization of capacity as a function of $x$ when the values of $y$ and $z$ are fixed at their optimal values: $y = y^*$ and $z = z^*$. This figure presents the distribution profiles of the Exponentiated Weibull (EW) fading of the UAV-to-OGS optical channel for various default parameters and its impact on UAV positioning. On the left, we can see from the illustration that a smaller tail probability corresponds to  more severe fading. The center graph illustrates the channel capacity of a UAV based AF relay. Evidently, under different levels of EW fading conditions, it is observed that the fading severity plays a pivotal role in determining the optimal UAV position. The right figure illustrates the relationship between channel capacity and the UAV's position  for a DF relay. For both AF and DF relays, we observe that as the fading levels becomes more severe, the optimal UAV location shifts closer to OGS in order to minimize the effect of fading in the optical channel.

\begin{figure*}
    \begin{minipage}[t]
    {\linewidth}
        \centering
        \includegraphics[width = 5.6in]{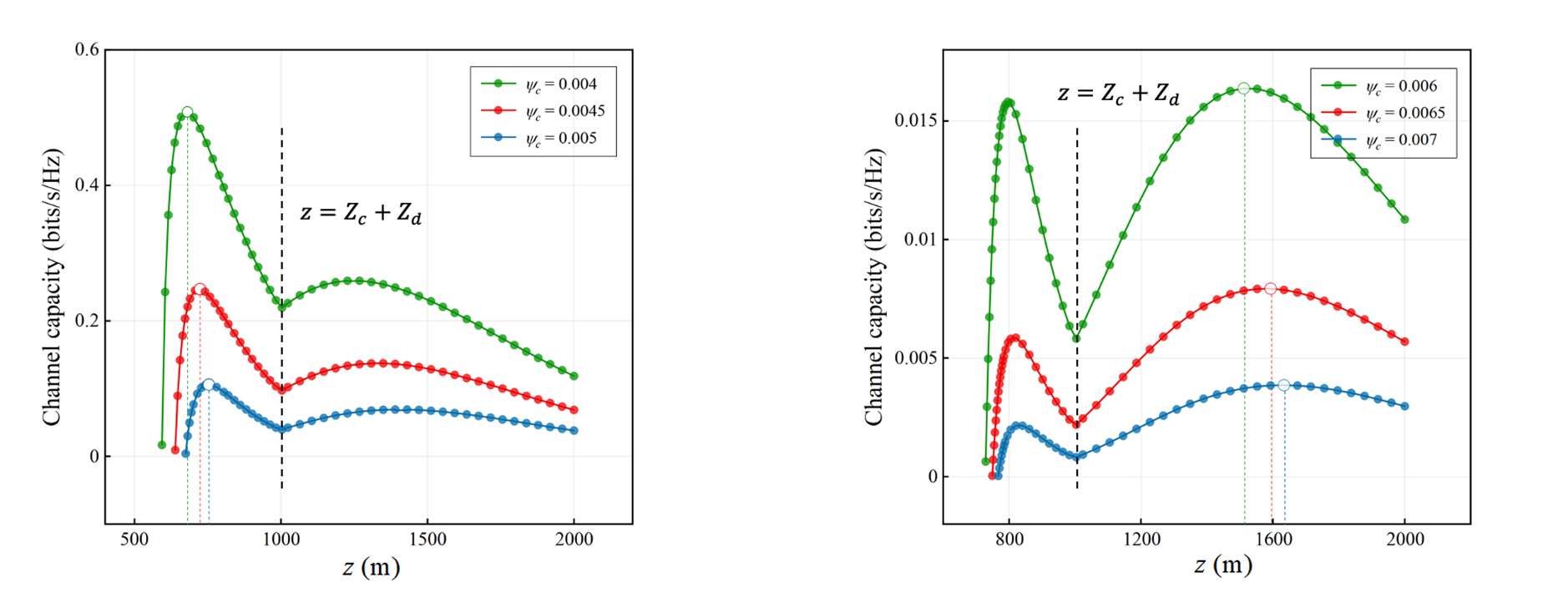}
        \vspace{-4mm}
        \caption{The figure show the channel capacity as a function of UAV's altitude for different cloud attenuation factor $\psi_c$.}
        \vspace{0mm}
        \label{fig:last 2}
    \end{minipage}
    \begin{minipage}[t]
    {\linewidth}
        \centering
        \includegraphics[width = 5.6in]{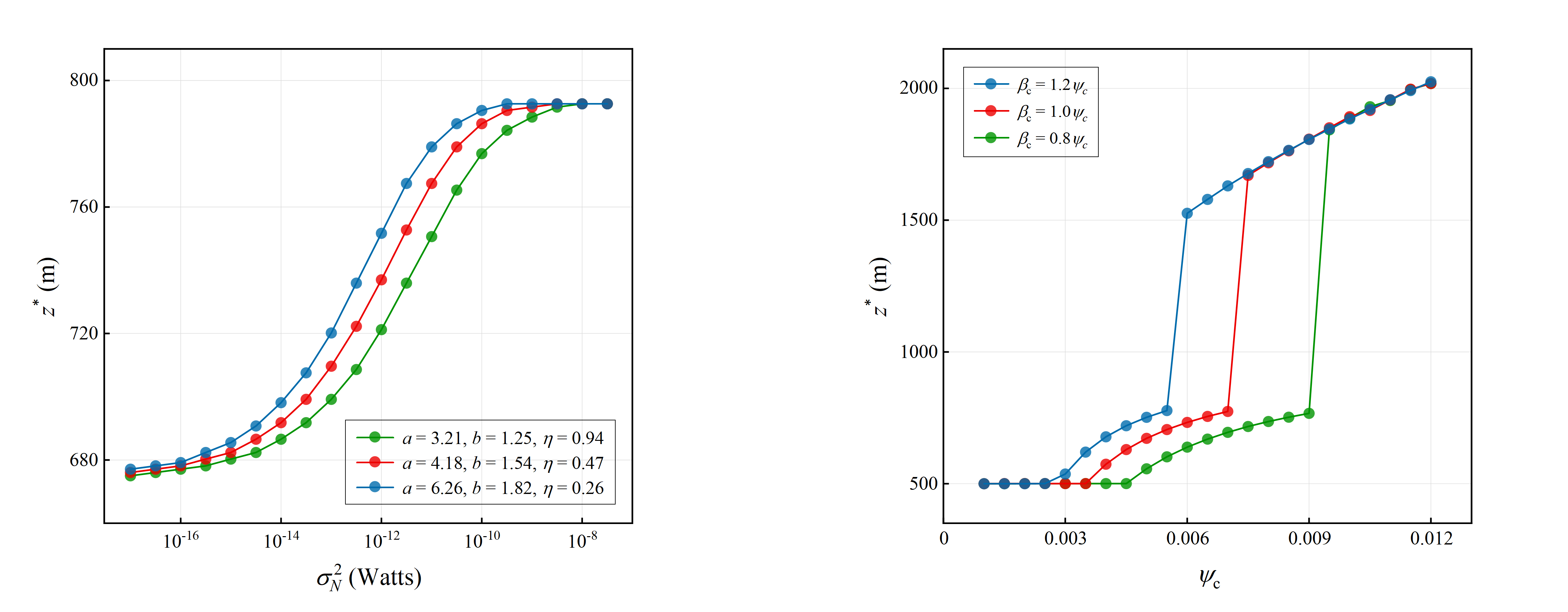}
        \vspace{-4mm}
        \caption{The figures show the optimal height of UAVs as functions of $\sigma_N^2$ (left) and $\psi_c$ (right) for different level fading channel (left) and different models of cloud attenuation factors (right).}
        \label{fig: final}
    \end{minipage}
\end{figure*}
\vspace{-1mm}
Fig.~\ref{OP3_new3} illustrates the impact of cloud attenuation factor $\psi_c$ and optical noise power $\sigma_N^2$ on the UAV optimal location $x^*$ for AF (left figure) and DF (right figure) relays. For one, this figure suggests an inverse relationship between $x^*$ and these channel parameters  due to the obvious reason that as the optical channel gets more impaired, the UAV has to shift closer to the OGS to improve the laser link (thereby minimizing $x^*$).  This figure also highlights the asymptotic regimes of the two relays discussed in Section~\ref{af_relay} and Section~\ref{df_relay}. When the conditions for asymptotic case 1 are true for the AF relay (high SNR in sensor-to-UAV link and low SNR in UAV-to-OGS link), we note that the relay settles above the OGS, i.e., the relay optimal position $x^* \to 0$ as $\sigma_N^2 \to \infty$. On the other hand, for asymptotic~case~3 (low SNR in both sensor-to-UAV and UAV-to-OGS links), the optimal position $x^*$ becomes independent of noise power $\sigma_N^2$ as $\sigma_N^2 \to \infty$. Moreover,  $\sigma_N^2 \to \infty \implies x^* \to \epsilon_0$ where $\epsilon_0 > 0$. This is in contrast with the DF relay case---shown in the right subfigure of Fig.~\ref{OP3_new3}---where the optimal position $x^* \to 0$ as $\sigma_N^2 \to 0$ irrespective of the SNR in the sensor-to-UAV link (asymptotic~case~1 for DF relay). For high SNR in UAV-to-OGS link, the optimal UAV position is given by the solution to \eqref{opt_prob_1} for both AF and DF relays. In this case, the optimal solution depends only on the sensor locations and their transmitted powers.






Next, we investigate the impact of the UAV's altitude on the end-to-end  capacity and explore how variations in the atmospheric environment affect the optimal altitude of the UAV. A critical consideration is that the energy $P_H(z)$---harvested by solar panels on the UAV---depends on the altitude of the UAV and the solar energy attenuation coefficient due to atmospheric cloud layer, $\beta_c$. Consequently, when the UAV is positioned within the cloud layer, that is, where $Z_d < z < Z_c + Z_d$, we not only need to account for the attenuation of laser signal energy transmitted by the UAV through the cloud layer, but  also the attenuation of solar energy as it passes through the cloud layer before being absorbed by the UAV solar panels. 

The interplay between the attenuation coefficients $\psi_c$ and $\beta_c$---where $\psi_c$ represents attenuation of laser signal through clouds and $\beta_c$ denotes  solar energy decay through clouds---is central to this optimization problem. It is crucial to emphasize that these parameters are inherently linked and are subject to simultaneous variations triggered by changing atmospheric conditions. However, these attenuation coefficients are strongly related to the wavelengths involved, and the wavelength-dependent attenuation is influenced by the cloud layer model which encompasses numerous variables. These include the size distribution of droplets within the cloud,  the complex refractive indices related to the corresponding wavelengths as well as the shape of the particles \cite{moll2007wavelength}. The relationship between $\psi_c$ and $\beta_c$, while crucial, is not the main focus of our study. Therefore, we have simplified the relationship between the two parameters into a linear model for our study in this paper.

Fig.~\ref{fig:last 2} shows the end-to-end channel capacity as a function of the UAV altitude under the assumption $\beta_c = \psi_c$, modulated by various laser signal energy attenuation factors $\psi_c$. For this figure as well as the remaining figures in this paper, the sensor powers are distributed as $\bm{P_m} = \begin{bmatrix}
        0.5, 1, 1.5, 2, 2.5, 3, 3.5, 4, 4.5, 5
    \end{bmatrix}$ Watts. Fig.~\ref{fig:last 2} distinctly showcases the bimodal nature of the capacity curves as a function of altitude $z$. The critical point---where the slope changes instantly from negative to positive---occurs when the UAV is at the uppermost boundary of the cloud layer at $z= Z_c + Z_d = 1000$ m. A major insight from this figure is the increase in optimal UAV altitude  as $\psi_c$ escalates. This implies that in the adverse atmospheric conditions, the UAV tends to ascend closer to the sun  to harvest more solar power to maintain the end-to-end link. A second major observation comes from the right subfigure in Fig.~\ref{fig:last 2}, where we note that as $\psi_c$ exceeds a particular value, the optimal altitude shifts abruptly from the first peak (optimal height inside the cloud) to the second peak (optimal height outside the cloud). To provide a comprehensive understanding of the relationship between changes in the UAV optimal altitude and the atmospheric cloud conditions, a deeper investigation is essential. 

In Fig. \ref{fig: final}, the left curve depicts the relationship between the optimal altitude of the UAV and optical thermal noise power $\sigma_N^2$. Here, we note that an increase in noise power---or conversely a decrease in the signal-to-noise ratio---prompts the UAV to fly higher to get closer to the sun in order to harvest more solar energy and reduce the impact of thermal noise on the performance. For large noise power, the optimal altitude becomes independent of the noise power due to \eqref{asymptot_AF_2}. Moreover, $\sigma_N^2 \to 0 \implies z^* \to 0$ due to \eqref{AF_asymptot2} and \eqref{signal_X}.

The right graph of Fig. \ref{fig: final} elucidates the impact of cloud channel coefficients $\psi_c$ and $\beta_c$  on the optimal height of UAV assuming a linear model between the two atmospheric parameters. We note that at minimal values of $\beta_c$ and $\psi_c$, the optimal altitude for UAV converges at the bottom of the cloud $z = Z_d$, a fact that is attributed to minimal solar energy attenuation by the clouds. However, below this altitude, the extinction coefficient of air---denoted as $\beta_a = 0.05$---significantly diminishes solar energy, making the cloud base an equilibrium point for UAV positioning. As $\beta_c$ and $\psi_c$ increase due to formation of clouds, the UAV's optimal altitude starts to rise gradually. A particularly notable observation is that for each assumed linear model, there is an abrupt transition in the UAV's optimal height from within the cloud layer to the region above the top boundary of the cloud as $\psi_c$ increases. This transition correlates with the dual peaks observed in Fig. \ref{fig:last 2}, where the optimal height shifts from the location of the first peak to the second in an abrupt fashion. Finally, as $\psi_c$ exceeds a certain threshold and continues to increase, the UAV's optimal altitude follows an approximately linear trajectory, with the three curves converging. This convergence is due to the fact that above the clouds, solar energy reaches the UAVs' solar panels without attenuation by the cloud, prompting the UAV to adopt similar altitude strategy across different atmospheric models.







\section{Conclusions}\label{conc}

In this paper, we present a comprehensive model for a solar-powered UAV-assisted RF/FSO link that is  aimed at facilitating data relay between distant sensors and an optical ground station (OGS). We demonstrate that there exists  an optimal UAV position in three dimensions at which the end-to-end capacity is maximized. Our analysis also highlights the impact of atmospheric conditions, such as cloud coverage, signal fading due to atmospheric turbulence and pointing error on the optical UAV-to-OGS link. A major finding of our study indicates that as the cloud channel becomes worse, the solar-powered UAV must be placed higher up in the sky to maximize the end-to-end rate. We believe that these findings will offer valuable insights for the design and implementation of future solar UAV-assisted RF/FSO communication links.




\bibliographystyle{IEEEtran}
\bibliography{ref.bib}

\end{document}